\documentclass[12pt]{article}

\usepackage[utf8]{inputenc}
\usepackage[T1]{fontenc}
\usepackage[backend=biber,style=apa]{biblatex}
\usepackage{graphicx}
\usepackage[svgnames]{xcolor}
\usepackage{tikz}
\usepackage{pgfplots}
\usepackage{pgfplotstable}
\usepackage{colortbl}
\usepackage{authblk}
\usepackage{booktabs}
\usepackage[hidelinks]{hyperref}
\usepackage{algorithm}
\usepackage{algpseudocode}
\usepackage{amsmath,amssymb}
\usepackage{acronym}
\usepackage{subcaption}
\usepackage{enumitem}
\usepackage{float}
\usepackage{forest}
\usepackage{fullpage}
\usepackage{setspace}
\usepackage{todonotes}
\usepackage{fancyhdr}

\allowdisplaybreaks

\usetikzlibrary{positioning,calc,arrows.meta,shapes,plotmarks}
\pgfdeclarelayer{foreground}
\pgfdeclarelayer{background}
\pgfdeclarelayer{secondBackground}
\pgfsetlayers{secondBackground,background,main,foreground}

\pgfplotsset{compat=1.17}
\usepgfplotslibrary{statistics,external}
\pgfplotstableset{
	/color cells/min/.initial=0,
	/color cells/max/.initial=1000,
	/color cells/textcolor/.initial=,
	color cells/.code={%
		\pgfqkeys{/color cells}{#1}%
		\pgfkeysalso{%
			postproc cell content/.code={%
				\begingroup
				\pgfkeysgetvalue{/pgfplots/table/@preprocessed cell content}\value
				\ifx\value\empty
				\endgroup
				\else
				\pgfmathfloatparsenumber{\value}%
				\pgfmathfloattofixed{\pgfmathresult}%
				\let\value=\pgfmathresult
				\pgfplotscolormapaccess
				[\pgfkeysvalueof{/color cells/min}:\pgfkeysvalueof{/color cells/max}]
				{\value}
				{\pgfkeysvalueof{/pgfplots/colormap name}}%
				\pgfkeysgetvalue{/pgfplots/table/@cell content}\typesetvalue
				\pgfkeysgetvalue{/color cells/textcolor}\textcolorvalue
				\toks0=\expandafter{\typesetvalue}%
				\xdef\temp{%
					\noexpand\pgfkeysalso{%
						@cell content={%
							\noexpand\cellcolor[rgb]{\pgfmathresult}%
							\noexpand\definecolor{mapped color}{rgb}{\pgfmathresult}%
							\ifx\textcolorvalue\empty
							\else
							\noexpand\color{\textcolorvalue}%
							\fi
							\the\toks0 %
						}%
					}%
				}%
				\endgroup
				\temp
				\fi
			}%
		}%
	}
}

\providecommand{\keywords}[1]
{
	\small	
	\textbf{\textit{Keywords---}} #1
}

\newtheorem{definition}{Definition}
\newcommand{\problemName}{Network Immunization Problem}

\newcommand{\solutionSet}{B}
\DeclareMathOperator*{\E}{\mathbb{E}}

\newcommand{\revised}[1]{\textcolor{black}{#1}}
\newenvironment{Revised}{\color{black}}{}

\newenvironment{SecondRevised}{\color{black}}{}

\AtEveryBibitem{\ifcategory{revisedBib}{\color{blue}}{}}
\DeclareBibliographyCategory{revisedBib}

\newcommand{\selfQuarantineProb}{p^Q_{\mathit{self}}}
\newcommand{\contactQuarProb}{p^Q_{\mathit{neighbor}}}
\newcommand{\notificationDelay}{t_{\mathit{notify}}^{\mathit{delay}}}
\newcommand{\maximumTracing}{t^{\mathit{max}}_{\mathit{trace}}}
\newcommand{\contactSpreadProb}{\beta_{\mathit{con}}}
\newcommand{\spontaneousSpreadProb}{\beta_{\mathit{spon}}}
\newcommand{\nrCloseContacts}{N_{\mathit{close}}}

\newcommand{\populationSize}{N}
\newcommand{\mutationRate}{\rho}
\newcommand{\tournamentSize}{\tau}
\newcommand{\nrNodesPerMeasure}{\nu}

\newcommand{\GAPopulation}{solution population}

\newcommand{\DTUEngTech}{DTU Eng.\@ Tech.\@}
\newcommand{\DTUManagement}{DTU Management}
\newcommand{\DTUFull}{full DTU}

\author[a]{Rowan Hoogervorst\thanks{Corresponding author}\thanks{A significant part of the work was carried out while the authors were working at the Technical University of Denmark.}}
\author[b]{Evelien van der Hurk\protect\footnotemark[2]}
\author[c]{David Pisinger}

\title{Simulation-Optimization Approaches for the Network Immunization Problem with Quarantining}
\date{}

\affil[a]{EDHEC Business School, 24 avenue Gustave Delory, 59057 Roubaix, France, rowan.hoogervorst@edhec.edu}
\affil[b]{CGI Netherlands, George Hintzenweg 89, 3068 AX Rotterdam, The Netherlands, evelien.vander.hurk@cgi.com}
\affil[c]{Department of Technology, Management and Economics, Technical University of Denmark, Akademivej 358, Kongens Lyngby, 2800, Denmark, dapi@dtu.dk}

\fancypagestyle{firstpage}
{
	\fancyhead[L]{}    
	\fancyhead[C]{DOI of published article: \href{https://doi.org/10.1016/j.ejor.2026.01.019}{\textcolor{blue}{10.1016/j.ejor.2026.01.019}}}
}

\begin{document}
	\maketitle
	\thispagestyle{firstpage}
	
	\begin{abstract}
		Vaccination has played an important role in preventing the spread of infectious diseases.
		However, the limited availability of vaccines and personnel at the roll-out of a new vaccine and the costs of vaccination campaigns often limit how many people can be vaccinated.
		Network immunization thus focuses on selecting a fixed-size subset of individuals to vaccinate so as to minimize the disease spread.
		In this paper, we consider simulation-optimization approaches for this selection problem.
		Here, the simulation of disease spread in an activity-based contact graph allows us to consider the effect of contact tracing and a limited willingness to test and quarantine.
		First, we develop a stochastic programming heuristic based on sampling infection forests from the simulation.
		Second, we propose a genetic algorithm tailored to the immunization problem that combines simulation runs of different sizes to balance the time needed to find promising solutions with the uncertainty resulting from simulation.
		Both approaches are tested on data from a major university in Denmark and disease characteristics representing those of COVID-19.
		Our results show that the proposed methods are competitive with a large number of centrality-based measures over a range of disease parameters and that especially the stochastic programming heuristic can outperform them for a considerable number of these instances.
		Finally, we compare network immunization against our previously proposed approach of limiting distinct contacts.
		Although, independently, network immunization has a larger impact in reducing disease spread, we show that the combination of both methods reduces the disease spread even further.
	\end{abstract}


	\section{Introduction}
	Immunization of a population through vaccination has been shown to play a vital role in reducing the spread of infectious diseases.
	Examples include the vaccination campaigns against smallpox, influenza, and, more recently, COVID-19.
	The latter has, e.g., reduced the health risks for individuals \autocite{tartof2021effectiveness,vasilei2021interim,mcnamara2022estimating} and allowed for a greater degree of opening up society \autocite{bauer2021relaxing,oliveramesa2022modelling}.
	However, immunization is often costly due to the cost of acquiring and administering vaccines.
	Moreover, time constraints and limited availability of vaccines during the roll-out of a new vaccine often make it necessary to prioritize some individuals for vaccination.
	Therefore, finding efficient immunization strategies that target the most influential individuals and achieve the greatest reduction in disease burden has shown to be an important topic of research.
	
	In this paper, we look at the immunization of a population that is represented through an activity-based contact hypergraph.
	Each individual is represented by a node in this graph, and hyperarcs represent planned activities that individuals are involved in over time.
	These activities could, e.g., be school classes, sports classes, or work meetings.
	During these activities, contacts occur between those individuals partaking in the activity, i.e., between some of the individuals that are part of the same hyperarc, allowing the disease to spread.
	The network immunization problem then focuses on selecting nodes from the graph to immunize, given a budget on how many individuals can be immunized.
	Our aim is hereby to limit the spread of the disease and thus to minimize the number of individuals that are infected over a given time horizon.
	
	Compared to the existing literature on network immunization, we consider a richer disease spreading model that considers the quarantining of infected individuals and the quarantining of exposed contacts as a result of contact tracing.
	Moreover, we take into account a limited willingness to test and quarantine, corresponding to the fact that not all individuals will opt to get tested and quarantine after becoming infected or being informed of an infected contact.	
	To measure the disease spread over the network under this disease spreading model with contact tracing, we rely on a simulation approach instead of direct graph measures.
	In particular, we use a simulation approach similar to the one in \textcite{bagger2022reducing}, which we  integrate into simulation-optimization methods for the network immunization problem.
	
	We consider an application focusing on the case of higher education for one of Denmark's largest universities, using data that was introduced in \textcite{bagger2022reducing}.
	Here, the population consists of students who would like to attend sessions for the classes they have subscribed to.
	The activities considered are thus scheduled course classes in which students meet and during which the disease can spread.
	Immunization in our setting corresponds to offering vaccination to individuals, representing a setting in which it would be possible to offer vaccines on an individual basis to a given number of students at the university.
	\revised{We note that our methods can also be used for vaccination in other settings in which contacts occur during scheduled activities, such as workplaces, sports clubs, hobby clubs, and primary and secondary schools.
	Moreover, they can be employed to determine interventions in high-risk environments in which it is possible to track and predict contacts to a reliable extent, such as in elderly health care institutions.
	}
	
	The contributions of this paper are fourfold.
	First, we define the network immunization problem with quarantining and contact tracing and describe how simulation can be used to determine the disease spread.
	Second, we propose a stochastic programming heuristic based on sampling infection forests from the simulation model.
	Third, we propose a parallelized genetic algorithm to solve the problem, which extensively makes use of the graph characteristics to find efficient solutions and combines small and large simulation runs.
	Fourth, we perform an extensive numerical study in which we show that our proposed methods are competitive with a large number of existing centrality measures and show the benefits of network immunization for our university application by comparing and contrasting to a scheduling policy that minimizes the number of distinct contacts \autocite{bagger2022reducing}.

	The paper is organized as follows.
	In \autoref{sec: literature review}, we discuss the relevant literature.
	In \autoref{sec: problem description}, we introduce the disease spreading model and formally define the considered network immunization problem.
	We propose both a stochastic programming heuristic and a genetic algorithm to solve the problem in \autoref{sec: genetic algorithm}.
	The data that we use from a major university in Denmark is described and analyzed in \autoref{sec: data}.
	We discuss the results obtained by our network immunization approach for this application in \autoref{sec: results}, where we  benchmark the proposed algorithms, evaluate the extent to which immunization can reduce the disease spread\revised{, and investigate the sensitivity of the proposed algorithms' performance to the disease parameters and their tuning parameters.}
	Lastly, the paper is concluded in \autoref{sec: conclusion}.
	
	\section{Literature Review}
	\label{sec: literature review}
	
	The study of network immunization policies for limiting disease spread is part of a larger stream of research on vaccine allocation, which considers the allocation of vaccines over, e.g., geographical, age, and social groups.
	See, e.g., \textcite{medlock2009optimizing},
	\textcite{yarmand2014optimal}, \textcite{enayati2020optimal},  and \textcite{lio2022optimizing}, who focus on the allocation of influenza and COVID-19 vaccines to different groups in the population, respectively.
	Network immunization problems characterize themselves by considering detailed contact networks, accounting for the fact that diseases tend to spread differently on realistic contact networks than in random graphs \autocite{satorras2001epidemic,newman2002spread}.
	Moreover, network immunization generally focuses on immunizing influential individuals in the population rather than on targeting groups as a whole based on, e.g., age or social characteristics.
	It should be noted that network immunization problems can also be found in other application areas, such as when looking at the spread of computer viruses \autocite{gao2011network}, the spread of harmful information in social networks \autocite{peng2019immunization}, \revised{the spread of a fire through a graph \autocite[the Firefighter Problem, see, e.g.,][]{finbow2009firefighter,hartnell1995firefighter}, and the spread of crises and bankruptcies in financial networks \autocite{garas2010worldwide,philippas2015insights}}.
	
	Traditionally, network immunization has focused on sequentially eliminating nodes from a network by ranking the nodes and choosing those nodes with the highest rank \autocite{pastor2002immunization}.
	The centrality of a node is a commonly used indicator of its importance in the network, and different types of centrality measures have been proposed.
	For example, \emph{degree centrality} looks at the number of neighbors adjacent to a node, where the assumption is that nodes with more neighbors are more likely to spread a disease \autocite{pastor2002immunization}.
	Another common centrality measure is \emph{betweenness centrality}, which looks at the number of times a given node is on the shortest path between any other two nodes \autocite{freeman1977set,anthonisse1971rush}.
	Other centrality concepts include those of \emph{eigenvector centrality} \autocite{bonacich1972factoring} and \emph{closeness centrality} \autocite{freeman1978centrality}.
	
	While the above measures are general indicators of a node's importance and not specific to preventing disease spread, an increasing number of papers are now focusing on measures that consider the specifics of disease spreading models.
	For example, it has been shown by \textcite{chakrabarti2008epidemic} that the epidemic threshold, i.e., \revised{the value of the effective spreading rate} below which the epidemic dies out, for \revised{an} SIS-epidemiological model equals the inverse of the largest eigenvalue of the adjacency matrix of the network.
	Therefore, multiple papers focus on maximizing the eigenvalue drop that is achieved by removing a node, where  \textcite{chen2016node} suggest approximating the eigenvalue drop using a so-called shield value and \textcite{mieghem2011decreasing} suggest different heuristic strategies for selecting a node.
	Another approach to better consider the disease spread dynamics is taken by \textcite{piraveenan2013percolation}, who suggest explicitly taking the current health state of each node into account when ranking the nodes, a measure they refer to as percolation centrality.
	
	Opposed to selecting nodes sequentially, which in general does not provide an optimal solution, authors have also looked at selecting the set of nodes to remove in an integrated way.
	\textcite{emmerich2020multiple} use quadratic optimization to find the set of nodes to immunize that minimizes the shield value and the cost of immunization.
	\textcite{saha2015approximation} propose approximation algorithms for maximizing the eigenvalue drop by either removing nodes or edges from the graph.
	Moreover, \textcite{nandi2016methods} propose algorithms for removing edges in a network to minimize \revised{different measures of its connectivity, and thus the expected spread of a disease through the network (note, though, that connectivity can be beneficial in other network immunization contexts)}.
	Another strategy used to minimize \revised{the graph's connectivity} is to disconnect the graph into multiple connected components.
	For example, \textcite{schneider2011suppressing} propose an algorithm to minimize the size of the largest connected components over the duration of the immunization process.
	Moreover, \textcite{ventresca2014randomized} propose a randomized rounding algorithm for finding the smallest possible subset of vertices to remove such that the graph is split into disconnected components of a given maximum cardinality.

	Among the approaches that try to find the set of nodes in an integrated way, multiple use common metaheuristics from the mathematical programming literature.
	For example, \textcite{deng2016optimal} use tabu search to minimize the size of the largest connected component of the graph after node removal.
	Especially relevant for this study, \textcite{maulana2017immunization} propose a genetic algorithm to maximize the drop in the largest eigenvalue after node removal.
	They benchmark their method against the algorithm presented by \textcite{chen2016node}, which is based on shield value, and show that better results can be found by means of their algorithm.
	
	Compared to the literature described above for the network immunization problem, we focus on a richer epidemiological model in this study.	
	In particular, we focus on a SEIR epidemiological model and take into account the effect of contact tracing and the willingness of people to test and quarantine.
	As a result, we do not focus on a direct graph metric in assessing the objective during optimization but use simulation to assess the effect of immunization.
	To the best of the authors' knowledge, this is the first study to look at optimization approaches for network immunization in such a rich epidemiological setting that requires simulation.

	\section{Problem Description}
	\label{sec: problem description}
	
	To model the spread of a disease through a population, we consider an activity-based contact hypergraph $G = (V,H,T)$ like in \textcite{bagger2022reducing}.
	Each node $v \in V$ in this graph represents an individual in the population that can be exposed to the disease.
	The hyperedges $h \in H$ indicate planned activities, thus connecting those individuals participating in the activity.
	Moreover, the set $T$ gives the time periods during which these activities take place, where $H_t \subseteq H$ denotes those activities that take place at time $t \in T$.
	An example of this hypergraph structure is depicted in \autoref{fig: hypergraph structure}, which shows 12 individuals represented by nodes and three planned activities represented by hyperarcs.
	Note, in particular, how each hyperarc connects all individuals participating in the activity.
	
	\begin{figure}[htbp]
		\centering
		
		\begin{tikzpicture}
			% Activity 1
			\foreach \a in {1,2,...,6}{
				\draw (\a*360/6: 2cm) node[draw,circle] (\a) {\a};
			}
			
			\draw[blue,dash dot] (1) -- (0,0) --  node[left=0.15cm] {$t = 1, 2, 6$} (2);
			\draw[blue,dash dot] (1) -- (0,0) --  (3);
			\draw[blue,dash dot] (1) -- (0,0) --  (4);
			\draw[blue,dash dot] (1) -- (0,0) --  (5);
			\draw[blue,dash dot] (1) -- (0,0) --  (6);
			
			% Activity 2
			\foreach \a in {7,8,...,12}{
				\draw (\a*360/6: 2cm) + (8,0) node[draw,circle] (\a) {\a};
			}
			
			\draw[red,dashed] (7) -- (8,0) --  (8);
			\draw[red,dashed] (7) -- (8,0) --  (9);
			\draw[red,dashed] (7) -- (8,0) --  (10);
			\draw[red,dashed] (7) -- (8,0) -- node[right=0.05cm] {$t = 1, 3, 5$}  (11);
			\draw[red,dashed] (7) -- (8,0) --  (12);
			
			% Activity 3
			\draw[Green,densely dotted] (8) -- (4,0) -- node[above right=0.15cm and -0.3cm] {$t = 1, 2, 6$} (1);
			\draw[Green,densely dotted] (8) -- (4,0) --  (6);
			\draw[Green,densely dotted] (8) -- (4,0) --  (5);
			\draw[Green,densely dotted] (8) -- (4,0) --  (10);
		\end{tikzpicture}
		
		\caption{A visualization of the activity-based contact hypergraph for an example in which 12 individuals (nodes) attend three activities (blue/dash-dotted, green/dotted, and red/dashed hyperarcs) spanning six time periods. The periods in which a hyperarc is active are denoted next to the hyperarc.}
		\label{fig: hypergraph structure}
	\end{figure}
	
	We use a compartmental model to describe the current health state of each individual in the population, i.e., each node in the graph. 
	In such a compartmental model, each individual is categorized to be in one of the compartments based on its current health state.
	We consider the following health states in our model:
	\begin{itemize}
		\item \textbf{S}: The individual is susceptible to the disease, i.e., can potentially become infected in the future.
		\item \textbf{E}: The individual has been exposed to the disease but is not yet infectious.
		\item \textbf{I}: The individual is infectious and can transmit the infection to other individuals.
		\item \textbf{R}: The individual has recovered from the disease.
	\end{itemize}
	
	Transitions between health states for each individual occur according to a discrete time Markov chain, where the probabilities to move between states are influenced by the contacts in the graph $G$.
	The possible transitions in our Markov chain are given in \autoref{fig:SEIR_model}.
	Here, an individual moves with probability $\beta_t$ from being susceptible (S) to being exposed (E) in time period $t \in T$ or stays susceptible with probability $1 - \beta_t$.
	The value of $\beta_t$ depends on the previous contacts that the individual had within the graph $G$.
	Individuals move with probability $\mu$ from state E to state I and with probability $\gamma$ from state I to state R.
	Note that state R is an absorbing state, meaning that we assume that recovered individuals become immune to the disease.
	This assumption is motivated by the relatively short time period that we consider in our simulation.

	\begin{figure}[htbp]
		\centering		
		
		 \begin{tikzpicture}
		 	\node[circle,draw,minimum size=1.75cm] (S) {$S$};
		 	\node[circle,draw,minimum size=1.75cm,right=of S] (E) {$E$};
		 	\node[circle,draw,minimum size=1.75cm,right=of E] (I) {$I$};
		 	\node[circle,draw,minimum size=1.75cm,right=of I] (R) {$R$};

			\draw[-{Stealth[length=2.1mm]}] (S) -- node[above] {$\beta_t$} (E);
			\draw[-{Stealth[length=2.1mm]}] (E) -- node[above] {$\mu$} (I);
			\draw[-{Stealth[length=2.1mm]}] (I) -- node[above] {$\gamma$} (R);
			
			\draw[-{Stealth[length=2.1mm]}] (S.north east) to[out=90,in=0] ($(S.north) + (0,0.6)$) node[above] {$1-\beta_t$} to[in=90,out=-180] (S.north west);
			\draw[-{Stealth[length=2.1mm]}] (E.north east) to[out=90,in=0] ($(E.north) + (0,0.6)$) node[above] {$1-\mu$} to[in=90,out=-180] (E.north west);
			\draw[-{Stealth[length=2.1mm]}] (I.north east) to[out=90,in=0] ($(I.north) + (0,0.6)$) node[above] {$1-\gamma$} to[in=90,out=-180] (I.north west);
			\draw[-{Stealth[length=2.1mm]}] (R.north east) to[out=90,in=0] ($(R.north) + (0,0.6)$) node[above] {$1$} to[in=90,out=-180] (R.north west);
		 \end{tikzpicture}		
		
		\caption{Transition rates for the discrete time Markov chain describing the health states of individuals at any timestep $t \in T$.}
		\label{fig:SEIR_model}
	\end{figure}
	
	We assume that the probability $\beta_t$ depends on two different components.
	First, infections can take place due to close contacts that the individual $v \in V$ had with infected individuals in the previous period, i.e., during the activities $\{h \in H_{t-1} \mid v \in h \}$.
	As such activities can be large in size, we assume that each individual has, on average, close contacts to $\nrCloseContacts$ others in each activity.
	\revised{For example, in our application for university education, $\nrCloseContacts$ reflects that students will only be close enough to a limited number of other students in large classes for infections to pass.}
	Each close contact \revised{that an individual had} with an infected individual has a probability of $\contactSpreadProb$ to spread the disease.
	Second, infection can occur due to contacts not explicitly modeled through the contact graph, i.e., contacts that occur outside of the modeled activities.
	In our \revised{university education application}, these spontaneous infections could, e.g., correspond to infections that occur outside of the university setting and from random encounters at the university that we did not model.
	We assume that such spontaneous infection occurs with probability $\spontaneousSpreadProb$ during each time period.
	Moreover, we assume that each individual also has a probability of $\spontaneousSpreadProb$ of being infected at the start of the time horizon.
	
	We also consider the effect of quarantining and contact tracing in the evaluation of the spread of the disease.
	Here, people go into quarantine after transitioning to the infectious state (I) with probability $\selfQuarantineProb$.
	Moreover, it is assumed that people are made aware of any close contacts they had with people who tested positive over the last $\maximumTracing - \notificationDelay{}$ days.
	Here, $\maximumTracing$ represents the maximum number of days during which contacts are traced, while $\notificationDelay$ represents the delay as a result of waiting to be tested, getting the result of the test, and communicating the result to close contacts.
	A close contact of an infectious person goes into quarantine with probability $\contactQuarProb$, which incorporates both the willingness of individuals to quarantine after being informed as well as the likelihood of contact tracing being successful.
	Persons who are in quarantine are unable to spread the disease, corresponding to removing them from all activities, i.e., hyperarcs, that occur during the time periods in which they are quarantined.
	
	Regarding the immunization of individuals, we assume that all selected individuals are immunized at the start of the time horizon.
	Moreover, we assume that immunization is fully immunity-inducing, meaning that immunized individuals are no longer susceptible to the disease.
	Considering these assumptions, immunized individuals will no longer play a role in the spread of the disease.
	While these assumptions seem strong in practice, they \revised{reflect that} we look at relatively short time horizons of weeks to months.
	Moreover, it should be noted that our simulation approach \revised{for determining the number of infections given some immunization decisions} is not dependent on these assumptions and can be easily extended to take both a partially effective vaccine and immunization throughout the time horizon into account.
	\revised{Hence, the genetic algorithm proposed later in this paper can also be easily extended to this more general setting as it directly evaluates solutions using this simulation model.}
	
	We can now formally define the network immunization problem.
	Let $f(G,X)$ be a random variable giving the number of infected individuals that is obtained over all time periods $T$ for some hypergraph $G = (V,H,T)$ and a set of immunized nodes $X \subseteq V$.
	The uncertainty underlying $f(G,X)$ comes from the transitions in our Markov chain, the potential quarantining of individuals, and the initial set of infected individuals, each of which are stochastic.
	Moreover, let $k$ be the immunization budget, i.e., the number of nodes in the graph $G$ that can be immunized.
	We can then state the problem as:
	\begin{definition}{(\problemName)}
		\label{def: problem definition}
		Select a set $X \subseteq V$ with $|X| \leq k$, such that the expected disease spread $\E \left[ f(G,X) \right]$ is minimized.
	\end{definition}
	As the dimensionality of the uncertainty space is large, and we are unaware of an exact method to efficiently estimate the expected number of infections when considering quarantining, we will use simulation to estimate $\E \left[f(G,X)\right]$.
	We thus focus on the development of \emph{simulation-optimization} methods to solve the problem.

	\section{Solution Methodology}
	\label{sec: genetic algorithm}
	In this section, we present both a stochastic programming heuristic and genetic algorithm to solve the \problemName{}.	

	\subsection{Stochastic Programming Heuristic}
	%Def Stoch Programming heuristic
 	The first solution approach that we consider is a stochastic programming heuristic.
	The \problemName{} can be seen as a two-stage stochastic programming problem in which the first stage defines \revised{the nodes to be immunized}, and the second stage determines the expected number of infections for a given set of immunized nodes. 
	In particular, consider the decision variable $x_v$ that defines whether node $v \in V$ is immunized.
	The stochastic programming problem can then be formulated as
	\begin{align}
		\min \; & \E \left[ f(G, \{v \in V \mid x_v = 1\}) \right] \label{eq: sp objective} \\
		\textbf{s.t.} & \sum_{v \in V} x_v \leq k, \label{eq: sp knapsack} \\
		& x_v \in \{0,1\} && \forall v \in V \label{eq: sp domain}.
	\end{align}
    The objective minimizes the expected number of infected nodes, while the constraint ensures that at most $k$ nodes are immunized. 
    Note that the feasibility of any solution $x$ is solely restricted by constraint \eqref{eq: sp knapsack} that enforces the immunization budget and is not affected by the second stage.
    \revised{This means that we consider a stochastic programming setting that resembles complete recourse, i.e., any set of immunized nodes chosen in the first stage is feasible for the second stage}.

    %Motivation for heuristic, that is not traditional SAA
    As we are unaware of an analytical representation that can efficiently determine the expected number of infected nodes, we propose to take a sample average approximation (SAA) approach. 
    SAA has proven to be a popular method to approximate the expectation in the objective of a stochastic programming problem and has been successfully used to solve a wide number of simulation-based optimization problems \autocite{kim2015guide}. 
    SAA works by taking the average realized objective over a set of samples. 
    However, as the number of infected nodes depends on a complex interaction between immunized nodes, activities, stochastic quarantining of infected individuals and neighbors, as well as the seed infections, using simulation to estimate the quality of a specific immunization strategy could easily require a sample defined by 100-200 simulation runs. 
    Consequently, a straightforward SAA approach would be intractable within, e.g., a branch-and-bound approach, as the evaluation of a single solution then requires multiple seconds in our simulation model and the feasible solution space is large (any subset of \revised{$k$} nodes).

    %General idea
    Therefore, we propose a SAA heuristic that uses infection forests from a sample of simulation runs without immunization, to next estimate the expected number of infected nodes by immunizing a subset of k nodes. 
    The main motivating idea is that the sampled infection forests reflect the infection dynamics within the proposed disease spreading model, and should thus give a good indication of the most influential nodes and the interactions that occur when immunizing them.

	%Definition of forests and trees     
	To formalize our approach, \revised{consider that we use $\sigma_p$ simulation runs.
	We then} denote by $\mathcal{F}_s = \{F_{s1},\dots,F_{sm}\}$ the forest representing the infections \revised{in simulation} run $s \in \{1,\dots,\sigma_p\}$.
	Each infection tree $F_{si}$ in the forest depicts the set of nodes, i.e., individuals, that are infected as the result of one initial infection. 
	Such a tree results from either infected nodes at the start of the simulation, or spontaneous (outside) infections occurring later on, where we denote $m$ as the sum of the total number of initial and spontaneous infections in the simulation run. 
	Together these trees define a forest, as nodes can at most be infected once. 
	Additionally, let $V_{\mathcal{F}_s}$ denote the nodes contained in forest $\mathcal{F}_s$. 
	An illustration of an infection forest is given in \autoref{fig: immunization illustration-1}, where nodes 1 and 6 are the initially infected nodes and both lead to a tree of further infections.

    %Impact of selecting nodes
    Immunizing a node in any of the trees in the infection forest corresponds to removing a sub-tree of nodes from the forest and thus prevents the infection of those individuals. 
    Namely, as \revised{we assume that} immunized nodes cannot be infected in our setting, they can also not pass on the infection to other individuals.
	This idea is further illustrated in \autoref{fig: immunization illustration-1}, where the immunization of node 7 leads to the nodes below it no longer being infected and the shown blue (dashed) sub-tree being deleted from the sampled forest.	
	As the infection trajectories are likely to be different in each simulation run, the impact of immunizing a node will differ between the sampled forests. 
	Thus, we want to select those nodes that minimize overall infections over a large sample of infection forests. 
	In our example in \autoref{fig: immunization illustration}, immunizing node 7 would be a good choice in both the first and second sampled infection forest, as it leads to the removal of a sub-tree consisting of 3 and 4 nodes\revised{,} respectively.
	\revised{On the contrary}, while immunizing node 13 looks promising based on the second sampled infection forest, immunizing this node has little impact in the first sampled infection forest due to it being a leaf node.

\begin{figure}[tb]
   \centering
	\begin{subfigure}[t]{0.47\textwidth}
		\centering
 		\begin{tikzpicture}
			\node[draw,rounded corners=0.5cm, fill=white] (front) at (0,0) {%
				\begin{forest}
					for tree={circle,draw,minimum size=1.5em, inner sep=1.25pt}
					[1
						[2]
						[3
							[4]
							[5]
						]
					]
				\end{forest}
				\quad
				\begin{forest}
					for tree={circle, draw, minimum size=1.5em, inner sep=1.25pt}
					[6
						[7, for tree = {blue,edge={blue,densely dashed}}
							[8
								[9]
								[10]	
							]
						]
						[11]
						[12
							[13]
							[14]
							[15]
						]
					]
				\end{forest}};
		\end{tikzpicture}
   		\caption{First sampled infection forest, with seed infections 1 and 6.}
		\label{fig: immunization illustration-1}
	\end{subfigure}
   ~ 
    \begin{subfigure}[t]{0.47\textwidth}
        \centering
        \begin{tikzpicture}
			\node[draw,rounded corners=0.5cm, fill=white] (front) at (0,0) {%
				\begin{forest}
					for tree={circle,draw,minimum size=1.5em, inner sep=1.25pt}
					[4
						[1
							[17]
						]
						[5
							[3]
							[19]
						]
					]
				\end{forest}
				\quad
				\begin{forest}
					for tree={circle, draw, minimum size=1.5em, inner sep=1.25pt}
					[12
						[7, for tree = {blue,edge={blue,densely dashed}}
							[8
								[9]
								[10]
								[20]
							]
						]
						[14]
						[13
							[6]
							[18]
							[15]
						]
					]
				\end{forest}
			};
		\end{tikzpicture}

		\caption{Second sampled infection forest, with seed infections 4 and 12.}
		\label{fig: immunization illustration-2}
    \end{subfigure}
 	\caption{Illustration of the effect of immunizing node 7 in two seperate infection forests.}
 	\label{fig: immunization illustration}
\end{figure}

	The problem to select a subset of $k$ nodes such that we remove a maximum of nodes from all sampled forests can be formulated as an Integer Programming (IP) problem. We introduce auxiliary variables $y_{vs}$ to denote whether individual $v \in V_{\mathcal{F}_s}$ is infected in forest $\mathcal{F}_s$ given the immunization decisions $x$.
	Let $P_{vs}$ denote all the nodes (excluding $v$ itself) that lie on the path from $v \in V$ to the root node of the tree in which $v$ is contained in sample $s \in \{1,\dots,\sigma_p\}$.
	We then obtain the following IP formulation:
	\begin{align}
		\min \; & \frac{1}{\sigma_p} \sum_{s = 1}^{\sigma_p}  \sum_{v \in V_{F_s}} y_{vs} \label{eq: obj} \\
		\text{s.t.} \; & \sum_{v \in V} x_v \leq k, \label{eq: immunization budget} \\ 
		& y_{vs} \geq 1 - x_v  - \sum_{v' \in P_{v s}} x_{v'} && \forall s \in \{1,\dots,\sigma_p\}, v \in V_{\mathcal{F}_s}, \label{eq: infection occurs} \\
		& x_v \in \{0,1\} && \forall v \in V, \label{eq: domain x}\\
		& y_{vs} \in \{0,1\} && \forall s \in \{1,\dots,\sigma_p\}, v \in V_{\mathcal{F}_s} \label{eq: domain y}.
	\end{align}
	The objective \eqref{eq: obj} minimizes the average number of infections that occur after immunizing the selected nodes in all infection forests, where the forests have equal weights. Thus, we assume the outcome of all simulation runs to have equal probability.
	Constraint \eqref{eq: immunization budget} ensures that the immunization budget is satisfied, i.e., that not too many individuals are immunized.
	Constraints \eqref{eq: infection occurs} determine if a node is infected given the set of immunized nodes.
	Here, the variable $y_{vs}$ can only become zero, i.e., $v$ is not infected, if node $v$ is immunized itself ($x_{v} = 1$) or if it is contained in a subtree of another node that is immunized ($\sum_{v' \in P_{v s}} x_{v'} \geq 1$).
	This corresponds precisely to what we earlier saw in \autoref{fig: immunization illustration}, where the immunization of a node leads to the deletion of the subtree in the infection forest below it.
	The remaining constraints \eqref{eq: domain x} -- \eqref{eq: domain y} enforce the domain of the decision variables.

    %Algorithm line out
    Our algorithm consists of the following steps. 
    First, we perform $\sigma_p$ simulation runs using the simulation model \revised{presented} in \autoref{sec: problem description} with no immunization of nodes. 
    Next, we solve formulation \eqref{eq: obj} -- \eqref{eq: domain y} based on the found infection forests using a commercially available IP solver.
	To speed up the solution process, we provide the solver with a starting solution that is determined based on the degree centrality measure \autocite{pastor2002immunization}.
	Here, the $k$ best variables are selected based on the degree centrality measure and the variables $x_v$ and $y_{vs}$ are set to the corresponding values in the starting solution. 
	Finally, the found set of immunized nodes is \revised{output}, and the quality of this solution can be determined by a new set of simulation runs.

    %heuristic explanation
    The above algorithm is a heuristic for the \problemName{} for two main reasons. 
    First of all, because the model \eqref{eq: obj} -- \eqref{eq: domain y} is based on a sample of simulation runs.
    Second, because the impact of immunizing specific nodes is estimated under the assumption that it will cancel all infections within the corresponding subtrees. 
    However, even under the same conditions as in the initial simulation run, nodes previously infected by the immunized node may still come in contact with other infected nodes later in the simulation and get infected after all.   
	Moreover, quarantine may also impact the time moments during which an individual engages in activities, meaning that nodes might become exposed to the disease at different time points.
	Hence, the number of infections determined in problem \eqref{eq: obj} -- \eqref{eq: domain y} only approximates the expected number of individuals as defined in \autoref{def: problem definition}. 
	Therefore, we will always evaluate the quality of the solution proposed by our heuristic through a new set of simulation runs that explicitly consider the set of immunized nodes.

	\subsection{Genetic Algorithm} 
	We additionally developed a metaheuristic approach based on a \emph{Genetic Algorithm (GA)} for the \problemName{}.
	A GA is a metaheuristic inspired by evolution that mimics the process of natural selection by modifying a \emph{population} of individual solutions \autocite{sivanandam2008introduction}.
	In particular, a GA typically combines existing solutions through crossover to find new solutions (offspring) and incorporates mutation to create diversity in the \GAPopulation{}.
	
	As we require simulation to evaluate the found solutions, it can take considerable time to evaluate the solutions in the \GAPopulation{}.
	Therefore, we developed an adapted GA framework in which we combine smaller and larger simulation runs.
	Here, the smaller simulations are used to quickly identify the most promising solutions from the \GAPopulation{}, while the larger simulation runs evaluate these solutions to get a better estimate of the true expected disease spread and thus reduce simulation variance.
	By additionally running the small simulations in parallel, a significantly larger number of iterations can be executed in this way.
	Our GA framework is illustrated in \autoref{fig: GA illustration}.
	In the remainder of this section, we explain the different components of our GA algorithm.
	
	\begin{figure}[htbp]
		\centering
		
		\begin{tikzpicture}
			\node[draw] (initial) {Determine initial \GAPopulation{}};	
			
			\coordinate[below=0.75cm of initial] (inbetween) {};
			
			% Evaluate solutions
			\node[draw, below left=0.95cm and 3cm of inbetween, align=left, red, text=black] (eval1) {Eval. \\ Sol.\ 1};
			\node[draw, below left=0.95cm and 1.5cm of inbetween,align=left, red, text=black] (eval2) {Eval. \\ Sol.\ 2};
			\node[draw, below=0.95cm of inbetween, red, text=black,align=left] (evalMid) {Eval. \\ Sol. $i$};
			\node[draw, below right=0.95cm and 1.7cm of inbetween, align=left, red, text=black] (evalN) {Eval. \\ Sol.\ $N$};
			\node[below left=1.35cm and 0.6cm of inbetween] {\dots};
			\node[below right=1.35cm and 0.8cm of inbetween] {\dots};
			
			\node[draw, below=1.05cm of evalMid, blue, text=black] (bestEval) {Evaluate best (candidate) solutions};
			
			\node[draw, diamond, aspect=1.65, below=0.7 of bestEval] (best) {New best sol.?};
			\node[draw, right=1.4cm of best] (update) {Update best found sol.};
			
			\node[draw, below=0.75cm of best] (crossover) {Apply crossover between the solutions};
			\node[draw, below=1cm of crossover] (mutate) {Apply mutation to the solutions};
			
			\node[draw, diamond, aspect=1.65, below=1cm of mutate] (stop) {Stop crit.\ met?};
			\node[draw, right=1.4cm of stop] (export) {Return best sol.};
			
			% Simulation descriptions
			\node[right=0.5cm of evalN,red] (small) {\small Small simulation ($\sigma_s$ runs)};
			\node[right=0.5cm of bestEval,blue] (large) {\small Large simulation ($\sigma_l$ runs)};
			
			% Connections
			\draw[-] (initial) -- (inbetween);
			
			\draw[->] (inbetween) -- (eval1.north);
			\draw[->] (inbetween) -- (eval2.north);
			\draw[->] (inbetween) -- (evalMid);
			\draw[->] (inbetween) -- (evalN.north);

			\draw[->] (eval1.south) -- (bestEval);
			\draw[->] (eval2.south) -- (bestEval);
			\draw[->] (evalMid) -- (bestEval);
			\draw[->] (evalN.south) -- (bestEval);			
			
			\draw[->] (bestEval) -- (best);
			
			\draw[->] (best) -- node[below] {Yes} (update);
			\draw[->] (best) -- node[right] {No} (crossover);
			\draw[->] (update) |- (crossover);
			
			\draw[->] (crossover) -- node[right] {New solutions} (mutate);
			\draw[->] (mutate) --  node[right] {Adapted solutions} (stop);

			\draw[-] (stop) -| ++(-4.8,2) node[near start, below] {No} |- (inbetween);
			\draw[->] (stop) -- node[below] {Yes} (export);
		\end{tikzpicture} 
		
		\caption{Illustration of our GA framework. \revised{Each} solution in the \GAPopulation{} is first evaluated in parallel using smaller simulation runs in each iteration, after which the most promising solutions are evaluated using a larger simulation run. Moreover, crossover and mutation are applied to obtain the \GAPopulation{} for the next iteration.}
		\label{fig: GA illustration}
	\end{figure}
	
	\paragraph{Representation of Solutions}
	Consider the contact hypergraph $G = (V,H,T)$ and an immunization budget allowing for the vaccination of $k$ nodes.
	Each solution $\solutionSet$ in our \GAPopulation{} then consists of $k$ genes, each of which represents a node $v \in V$ selected for immunization in solution $\solutionSet$.
	As nodes with (very) low centrality are unlikely to be good candidates for immunization, we limit the search space to nodes with a high centrality on at least one of several centrality measures.
	For each of these considered centrality measures, denoted by the set $M$, we calculate the centrality score for each node at the start of the algorithm.
	Let $V_m \subseteq V$ be the set of nodes selected according to centrality measure $m \in M$, where $V_m$ contains the $\nrNodesPerMeasure < |V|$ nodes with highest centrality score on measure $m$.
	Each solution in the \GAPopulation{} is then of the form
	\begin{equation}
		\solutionSet{} = \{v_1, v_2,\dots,v_k\} \quad \mbox{ where } v_i \in \bigcup_{m \in M} V_m,
	\end{equation}
	meaning that only those genes are considered that belong to the $\nrNodesPerMeasure$ nodes with highest centrality score for at least one of the centrality measures.
	Note that $\nrNodesPerMeasure$ should be chosen such that $| \bigcup_{m \in M} V_m | \geq k$ and that we normally would like to choose $\nrNodesPerMeasure$ significantly larger than $k$ to find a good balance between the increased performance that can be expected from searching a smaller solution space and the potential reduction in solution quality.
	
	\paragraph{Initial \GAPopulation{}}
	In each iteration of the algorithm, we consider a \GAPopulation{} consisting of $N$ solutions.
	At the start of the algorithm, an initial population is generated in which a few solutions are selected based on the considered centrality measures $M$, while the other solutions are randomly selected.
	Here, we add a solution for each centrality measure $m \in M$ and let the genes of this solution be the  first $k$ nodes in the ranking provided by that centrality measure.
	In this way, we ensure that there are solutions in the initial \GAPopulation{} that likely lead to a low disease spread.
	The remaining $N - \vert M \vert$ solutions are then chosen randomly from the search space $\cup_{m \in M} V_m$ to ensure a diverse initial \GAPopulation{}.

	\paragraph{\revised{Evaluation of Solutions}}
	Each individual solution is evaluated \revised{using the simulation model} described in \autoref{sec: problem description}.
	As we use simulation to estimate the number of expected infections in the SEIR model for a particular contact graph, we consider the average \revised{number of infections over} all simulation runs.
	Therefore, \revised{the evaluation score} for a solution $\solutionSet$ is
	\begin{equation}\label{eq:fitnessScore}
		\mathit{\revised{score}}(\solutionSet) = \frac{\sum_{s = 1}^{\sigma} \tilde{f_s}(G,\solutionSet)}{\sigma},
	\end{equation}
	where $\tilde{f_s}(G,\solutionSet)$ denotes the number of infections in simulation run $s \in \{1,\dots,\sigma\}$ for the immunized nodes $\solutionSet$.
	The number of simulation runs depends on the phase of the genetic algorithm, as illustrated in \autoref{fig: GA illustration}.
	In particular, we first select, in each iteration of the GA, the $\lambda \in \mathbb{N}^+$ most promising solutions from our \GAPopulation{} based on computing the \revised{evaluation score} using a smaller number of simulation runs $\sigma_s$.
	For these $\lambda$ most promising solutions, a more \revised{accurate evaluation score} is then computed using $\sigma_l$ simulation runs to determine if any of these solutions improves on the incumbent solution.
	Here, it holds that $\sigma_l \gg \sigma_s$.
	In conclusion, the \revised{aim} in our GA is to find a solution $\solutionSet^* \subseteq \bigcup_{m \in M} V_m$ that minimizes the expected number of infections, i.e., 
	\begin{equation}\label{eq:fitnessScore-conclusion}
		\solutionSet^* = \underset{\solutionSet \subseteq \bigcup_{m \in M} V_m, |\solutionSet|=k}{\arg\min} \mathit{\revised{score}}(\solutionSet).
	\end{equation}
	
	\paragraph{Selection}
	The selection process includes elitist selection, where the best $\epsilon$ individual solutions from the current generation are directly moved to the next generation without crossover or mutation. 
	This ensures that good solutions remain present throughout the search process.
	The remaining $N - \epsilon$ solutions are generated by breeding pairs of solutions from the current generation, \revised{where each parent solution is chosen} through tournament selection.
	In such a tournament selection, $\tournamentSize$ solutions are randomly selected from the \GAPopulation{} and the one among them with the best \revised{evaluation score} is \revised{selected}.
	
	\paragraph{Crossover and Mutation}
	In each iteration, every pair of parent solutions, \revised{as selected by tournament selection}, produces two children solutions. 
	The mating process uses an adaptation of uniform crossover. 
	Here, genes that are included in both parents are first assigned to be part of both children.
	The other nodes for each child are then uniformly selected from the remaining genes of each parent, meaning that each gene has an equal chance of ending in any of the two children.
	Therefore, the approach prevents duplicates, i.e., a child cannot have two of the same genes.
	
	To ensure genetic diversity, mutation occurs during each generation.
	Here, each gene is mutated with probability $\rho$, leading to $\rho k$ genes, on average, being mutated per solution in each iteration.
	When a gene is mutated, a random node is selected uniformly from the set $\cup_{m \in M} V_m$ in such a way that the new gene is not already present in the solution.
	
	\paragraph{Stopping Criterion}
	We use a time-based stopping criterion for the genetic algorithm, meaning that the genetic algorithm is continued until a certain wall clock time limit  is reached.
	Note that this implies that the number of iterations depends on the instance and computing infrastructure.
	
	\section{Data}
	\label{sec: data}
	
	To test the proposed methods, we use course data from the \emph{Technical University of Denmark (DTU)}, one of the 8 major universities in Denmark.
	This dataset was first introduced by \textcite{bagger2022reducing}.
	The course data describes the classes that students have subscribed to for the fall semester of 2020 and the teaching sessions, such as lectures and exercise classes, that have been planned for these classes.
	In total, the data describes the preferences of over 9500 individual students who subscribed to about 650 courses.
	On average, each student takes 3--4 classes, leading to about 34500 individual course subscriptions.
	Moreover, most courses have one or more sessions each week, meaning that the total number of contacts over the whole 13-week semester is higher.
	
\begin{figure}[htbp]
		\centering
		\includegraphics[scale=0.35]{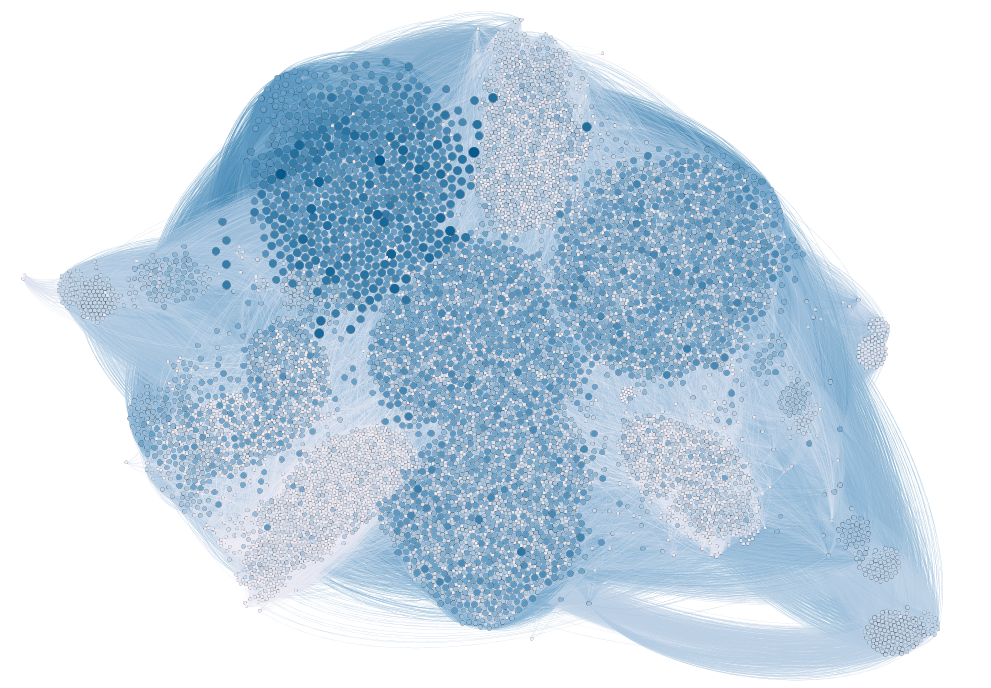}
		\caption{Visualization of the DTU contact graph $G'$, where each node indicates a student and an arc connects two students if they participate at least once in the same course. The node size and color intensity indicate the connectivity of a node, \revised{i.e., more connected nodes are darker and larger.}}
		\label{fig:student-graph}
	\end{figure}

	The course subscriptions of students lead to a contact hypergraph $G$, where each hyperarc represents a course session being attended by a certain group of students.
	As each hyperarc connects all individuals in the activity, each hyperarc can also be represented as a complete sub-graph connecting these individuals, which provides a regular graph $G'$.
	In this graph, an arc is present between any two students if they are participating together in at least one course session, meaning that we aggregate over the different time periods $T$.
	This graph $G'$ is depicted in \autoref{fig:student-graph}, where nodes are colored according to the number of contacts they have.
	Moreover, summary statistics for this graph $G'$ are given in the upper part of \autoref{tab:studentGraph}.
	
	The results in \autoref{tab:studentGraph} show that the average degree is large, which is explained by the fact that students are connected with all other students in the classes they attend.
	Moreover, both the diameter and the average shortest path length are low, indicating that a disease can generally spread rapidly in the network.
	Together, these properties show that the studied network exhibits many properties of a small world \autocite{watts1998collective}.
	It should be noted, though, that the connectivity in the simulation is limited by the number of close contacts per activity $\nrCloseContacts$.
	However, as course sessions are generally repeated on a weekly basis, students may still become a close contact to many of the other students in the class over the semester as they change seating over these different sessions.
	
	\begin{table}[htbp]
		\centering
		
		\caption{Graph structure properties of the \DTUFull{} contact graph $G'$ and the smaller department graphs, where $V$ denotes the number of nodes and $E$ the number of edges. $\bar{\delta}$ represents the average degree, $D$ the network's diameter, $L$ the average shortest path length, and $C$ the average clustering coefficient. \revised{$L_{ER}$} and \revised{$C_{ER}$} are calculated for an Erdős–Rényi random graph with the same \revised{number} of nodes and edges.}
		
		\begin{tabular}{@{}lcccccccc@{}}
			\toprule
			Graph & $|V|$ & $|E|$ & $\bar{\delta}$ & $D$ & $L$ & $C$ & $\revised{L_{ER}}$ & $\revised{C_{ER}}$ \\ \midrule
			DTU Full & 9,602 & 1,668,884 & 347.61 & 5 & 2.44  & 0.64 & 1.96 & 0.036 \\
			\midrule
			\DTUManagement{} & 2,415 & 341,511 & 282.82 & 4 & 2.09 & 0.84 & 1.88 & 0.117 \\ 
			\revised{\DTUEngTech{}} & 1,933 & 134,703 & 139.37 & 7 & 2.76 & 0.88 & 1.93 & 0.072 \\
			\bottomrule
		\end{tabular}
		
		\label{tab:studentGraph}
	\end{table}

	As the visualization of $G'$ in \autoref{fig:student-graph} seems to indicate the presence of some highly connected communities, we additionally evaluated the community structure of the graph $G'$.
	We used the Louvain method \autocite{blondel2008fast} for this, which was found to be efficient for both synthetic and real-world networks \autocite{lancichinetti2009community,yang2016comparative}.
	The results of the Louvain method on the DTU contact graph are given in \autoref{tab: community structures}.
	Here, we also estimate a mixing coefficient $\mu$, which is the ratio of a node's external neighbors, i.e., neighbors that are in a different community, to the total degree of the node \autocite{lancichinetti2009community}. 
	The modularity score of 0.558 shows that $G'$ has a moderate to strong community structure, where the Louvain method identifies 9 different communities.
	The mixing parameter $\mu$ further confirms these findings, as it shows that, on average, less than 25\% of the edges connected to a node \revised{lead} to a node outside of the node's own community.
	
	\begin{table}[htbp]
		\caption{Topological community properties of the DTU contact graph $G'$.}
		\centering
		\begin{tabular}{lr} 
			\toprule
              	Topological Property & Value         \\
				\midrule
				Number of communities $C$& 9 \\
				Modularity $Q$ & 0.558 \\
				Estimated mixing coefficient $\mu$ & 0.242 \\
				\bottomrule
		\end{tabular}
		\label{tab: community structures}
	\end{table}
	
	\subsection{Department Subgraphs}
	As we would like to evaluate our methods on graphs with varying degrees of connectivity, we created additional (smaller) instances by considering subgraphs that cover single departments at DTU: DTU Management and DTU Engineering Technology (\revised{\DTUEngTech{}}).
	We then created a contact graph by considering all students taking courses at this department and only those classes that are offered by this department.
	Note that DTU students generally take courses from multiple departments but often follow the majority of their courses at one to two departments.
	
	Summary statistics for these two additional contact graphs are given in the lower part of \autoref{tab:studentGraph}.
	It can be seen that both graphs are significantly smaller than the \DTUFull{} contact graph, having about 20--25\% of the number of nodes and about 8--20\% of the edges of the full contact graph.
	It can also be seen that the average number of neighbors is slightly lower, which is not surprising considering that we are looking at (edge-induced) subgraphs.
	Instead, one can see that the average clustering coefficient is higher for these subgraphs, indicating that the nodes tend to be more clustered together.
	The two new graphs differ in the diameter and average path length, where the graph for \DTUManagement{} has a slightly lower path length and diameter than the full graph and the \revised{\DTUEngTech{}} graph a higher one.
	Based on the above, it can be concluded that these graphs are more clustered than the full graph, where the \DTUManagement{} graph additionally shows to be very connected.
	
	\subsection{Disease Characteristics}
	The disease characteristics that we consider are based on those of COVID-19.
	In particular, the chosen values are mostly based on the data provided for the Danish society by the \emph{Statens Serum Institute (SSI)}, which is under the auspices of the Danish Ministry of Health.
	An overview of the used parameter values is given in \autoref{table:seir-model-param}.
	
	\begin{table}[htbp]
		\centering
		\caption{The SEIR model parameters used for simulations. Some parameters take a fixed value over all simulations, while others are varied between a minimum and maximum value.}
		\begin{tabular}{lrrr}
			\toprule
			Parameter & Minimum & Maximum & Value \\
			\midrule
			$\selfQuarantineProb$ & -- & -- & 0.5 \\
			$\contactQuarProb$ & -- & -- & 0.4 \\
			$\mu$ & -- & -- & $\frac{1}{4}$ \\
			$\gamma$ & -- & -- & $\frac{1}{6}$ \\
			$\maximumTracing$ & -- & -- & 14 \\
			$\notificationDelay$ & -- & -- & 2 \\
			$\nrCloseContacts$ & -- & -- & 10 \\
			$\spontaneousSpreadProb$ & -- & -- & 0.0003 \\
			$\contactSpreadProb$ & 0.15 & 0.35 & -- \\
			\bottomrule
		\end{tabular}
		\label{table:seir-model-param}
	\end{table}

	The values of all parameters except $\contactSpreadProb$ are the same over the different experiments.
	The values of $\mu, \gamma, \maximumTracing, \notificationDelay$ and $\nrCloseContacts$ equal those used in \textcite{bagger2022reducing}, which further motivates their choice.
	Moreover, the values for the probability of self-quarantining $\selfQuarantineProb$ and $\contactQuarProb$ are based on earlier experiments in \textcite{bagger2022reducing}, who showed that these are values in which quarantining has a clear impact and which are also not unrealistically high.
	Note in particular that values close to 1 have shown to be unrealistic in practice, see, e.g., \textcite{davis2021contact}.
	In addition, the value of $\spontaneousSpreadProb$ is chosen as the middle value of the range for this parameter used in \textcite{bagger2022reducing}.
	The value of $\contactSpreadProb$ differs over the runs, \revised{as this parameter} will, in practice, depend strongly on the characteristics of both the activity and the room in which the activity takes place.
	The chosen range for this parameter is roughly based on the values used within the national models developed by SSI \autocite{statens2020teknisk}.
	\revised{Moreover, next to varying the value of $\contactSpreadProb$ over the runs in our experiments, we will also consider a setting in \autoref{sec: varying contact rate} where the value of $\contactSpreadProb$ is dependent on the activity in which the close contact takes place and thus varies between the hyperarcs $H$ in the graph.
	In addition, we evaluate the performance of the solutions found by our proposed immunization methods for a wider range of disease parameters in \autoref{sec: robustness}.
	} 
	
	\section{Numerical Study}
	\label{sec: results}
	
	In this section, we evaluate the proposed solution methods numerically for the DTU contact graph by looking at the resulting number of infections.
	Our aim is two-fold.
	On the one hand, we would like to investigate the benefit that the stochastic programming heuristic and genetic algorithm provide compared to using existing graph-based measures.
	On the other hand, we would like to evaluate how strong the effect of network immunization is for the given contact graphs at different immunization rates \revised{and how this effect is impacted by the disease parameters}.
	In the remainder of this section, we first introduce the benchmark methods and the setup of our experiments.
	Afterwards, we discuss the performance of the immunization methods, analyze the extent to which their solutions coincide, \revised{and investigate the robustness of the found solutions to changes in the disease parameters}.
	Moreover, we analyze the impact of network immunization by comparing its results to that of the scheduling policy introduced in \textcite{bagger2022reducing} that minimizes the number of distinct contacts.
	
	\subsection{Benchmark Methods}\label{sec:sequential}
	
	As illustrated by our literature review, a large number of graph-based methods have been proposed for solving network immunization problems.
	We will use these to benchmark our proposed solution methods, as well as within the genetic algorithm to determine the search space and the initial solution.
	We consider the following benchmark methods in our numerical study:
	\begin{enumerate}[label=M\arabic*]
		\item \textit{Random}: The nodes to immunize are chosen uniformly at random. We will consider the best (in-sample) solution out of 10 randomly generated solutions.
		\item \label{meth: degree} \textit{Degree centrality (Degree)}: Nodes are ranked according to the number of nodes adjacent to them, i.e., their number of neighbors. Nodes with more neighbors are then prioritized for immunization \autocite{pastor2002immunization}.
		\item \label{meth: harmonic} \textit{Harmonic centrality}: A centrality measure that looks at the path distance of a node to the other nodes in the graph \autocite{rochat2009closeness}. Nodes with a shorter distance to the other nodes are seen as more central and thus prioritized for immunization. This measure is similar to closeness centrality \autocite{freeman1978centrality}, but unlike closeness centrality, it also applies to disconnected graphs.
		\item \label{meth: eigenvector} \textit{Eigenvector centrality}: A centrality measure proposed by \textcite{bonacich1972factoring}  based on the idea that a node's importance is related to its neighbors' importance. It can be computed by determining the principal eigenvector of the adjacency matrix, i.e., the eigenvector corresponding to the largest eigenvalue.
		\item \label{meth: betweenness} \textit{Betweenness centrality}: A centrality measure that was proposed by \textcite{freeman1977set}, which looks at the number of times a node lies on the shortest path between any other nodes. 
		\item \textit{Community Bridge Finder (CBF)}: A community-based algorithm proposed by \textcite{salathe2010dynamics} that aims to find bridge nodes, i.e., nodes that connect different communities. The found bridge nodes are prioritized for immunization.
	\end{enumerate}	
	It should be noted that the above methods are defined for general graphs and not for the activity-based contact hypergraph $G$ considered in this study.
	Therefore, we apply these methods for the graph $G'$ considered before, in which each hyperarc in the hypergraph is replaced by the complete graph between all nodes that are part of the activity.
	
	An additional consideration for the centrality-based measures (\ref{meth: degree} -- \ref{meth: betweenness}) is that they can be applied both in a \emph{static} and \emph{dynamic} way.
	In the static approach, the score for all nodes is computed once and the $k$ nodes with the highest nodes are then selected to be immunized.
	In the dynamic approach, the centrality score of all remaining nodes is re-computed after each node removal, and only the (remaining) node with the highest score is removed in each iteration.
	Due to the computation time of the different methods, we will use a dynamic approach for the Degree centrality (\ref{meth: degree}) measure in our experiments and a static approach for the other centrality measures.
	
	\subsection{Experimental Setup}
	To evaluate the solutions of our methods and the benchmark methods, we will look at the total number of contact infections (CI).
	We have chosen here to evaluate the number of contact infections over the number of total infections, as the immunization strategy only has a direct effect on the contact infections.
	Instead, spontaneous infections can only be prevented if people are in quarantine, meaning that a policy leading to many infections might actually lead to a lower number of spontaneous infections due to more people being in quarantine.
	All evaluations were made out-of-sample, i.e., using a different stream of random numbers for the simulation than used within the simulation-optimization methods themselves, and are based on 200 simulation runs.
	
	The parameters used for the genetic algorithm in our experiments are given in \autoref{tab:ga-parameterTable}.
	Here, the number of highest ranking nodes $\nrNodesPerMeasure$ selected in the solution representation and the mutation rate $\rho$ are chosen depending on the number of nodes in the graph and the number of nodes to be immunized, respectively.
	Fixed values are chosen for the other parameters.
	Note that we choose the number of simulation runs $\sigma_s$ to identify promising solutions significantly lower than the number of simulation runs $\sigma_l$ to evaluate the best-found solutions.
	Moreover, a relatively small tournament size $\epsilon$ is selected to ensure that also slightly lower-scoring parents are chosen for crossover, especially as nodes are already chosen based on their centrality score within the solution representation.
	In addition, note that we use \revised{multiple centrality-based} measures discussed in \autoref{sec:sequential} to define the solution representation and select some of the initial solutions.
	Furthermore, we use one additional centrality measure that is specific to our setting of university education:
	\begin{enumerate}[label=M\arabic*,resume]
			\item \label{meth: neighbor} \textit{Neighboring weights}: Considers a weighted network in which the edge weights are computed based on the number of courses a pair of students attend together. The importance of a node is then determined by the sum of the neighboring edge weights, where nodes with higher importance are prioritized for immunization. 
	\end{enumerate}
	\revised{The impact of the evolutionary parameters $\populationSize, \mutationRate, \tournamentSize,$ and $\epsilon$ on the performance of the GA is further studied in \autoref{sec: GA tuning}.}
	
	The only parameter that needs to be specified for the stochastic programming heuristic is the parameter $\sigma_p$ that determines the number of infection forests.
	Here, we use $\sigma_p = 300$ simulation runs for the full contact graph to generate the infection forests to balance the quality of the solutions with the computation time of the solved IP problems.
	We use a larger number of runs $\sigma_p = 400$ for the department contact graphs, considering their smaller size and the better solution quality that is to be expected for a larger number of samples.
	The effect of the parameter $\sigma_p$ will be further studied in \autoref{sec: SP samples}.
	
	All experiments were performed on an Intel Xeon Gold 6142 processor, utilizing 8 CPU cores and \revised{a maximum of 60GB} of internal memory.
	Each immunization method was given a maximum computation time of three hours, translating to a time-based cut-off of three hours for the genetic algorithm and a maximum IP computation time of three hours for the stochastic programming heuristic.
	The IP model in the stochastic programming heuristic was solved using the Gurobi 10.0 solver \revised{with the allowed optimality gap for a solution set to 0.5\%.}
	The simulation model for determining the disease spread and all immunization methods were programmed in the Java programming language.
	
	\begin{table}[htbp]
		\begin{tabular}{cll}
				\toprule
				Parameter	& Description & Value \\ 
				\midrule
				\multicolumn{1}{c}{N} & Size of \GAPopulation{} & 50 \\
				\multicolumn{1}{c}{$M$} & Used centrality measures & $\{\textrm{M2}, \textrm{M3},\textrm{M4},\textrm{M7}\}$ \\
				$\nrNodesPerMeasure$ & Number of highest-ranking nodes per centrality measure & $\frac{\mid V \mid}{2}$ \\
				\multicolumn{1}{c}{$\tournamentSize$} & Tournament size of the tournament selection & 4\\
				\multicolumn{1}{c}{$\epsilon$} & Number of solutions chosen in elitist selection & 5 \\
				\multicolumn{1}{c}{$\mutationRate$} & Average mutation rate for each individual & 0.05 \\
				\multicolumn{1}{c}{$\sigma_s$} & Number of simulation runs in small simulations & 25 \\
				\multicolumn{1}{c}{$\sigma_l$} & Number of simulation runs in large simulations & 150 \\
				\multicolumn{1}{c}{$\lambda$} & Number of promising solutions evaluated per iteration & 3 \\
				\bottomrule
		\end{tabular}
		\caption{Overview of parameters used in the GA.}
		\label{tab:ga-parameterTable}
	\end{table}

	\subsection{Comparison of Immunization Approaches}
	We now compare the stochastic programming heuristic and genetic algorithm to the benchmark methods (M1) -- (M6).	
	Here, we explored the performance of the immunization methods both for different immunization rates, i.e., values of $k$, and contact infection probabilities $\contactSpreadProb$.
	\autoref{fig: performance over immunization budgets} shows the results of the immunization methods for varying immunization rates (10\%, 20\% and 30\% of the population size) for a fixed contact infection probability of $\contactSpreadProb{} =0.25$.
	\autoref{fig: performance over CI} shows the results of the methods for varying contact infection probabilities ($\contactSpreadProb{} \in \{0.15,0.25,0.35\}$) for a fixed immunization rate of 20\%.
	In both of these figures, the results correspond to the average number of contact infections after immunization, obtained by means of the simulation model discussed in \autoref{sec: problem description}.
	
	\pgfplotscreateplotcyclelist{methodsList} {
		{black,mark=*},
		{red,every mark/.append style={fill=red!80!black}, mark=square*},
		{brown!60!black,every mark/.append style={fill=brown!80!black},mark=otimes*},
		{densely dashed,brown!60!black,every mark/.append style={solid,fill=brown!80!black},mark=pentagon*},
		{yellow!60!black,mark=star},
		{densely dashed,yellow!60!black,every mark/.append style={solid},mark=square*},
		{orange,every mark/.append style={solid},mark=halfcircle*},
		{dashed,blue,every mark/.append style={solid},mark=halfsquare*},
		{blue,mark=diamond*}
	}
	
	\begin{figure}[htbp]
		\begin{subfigure}{0.49\textwidth}
			\begin{tikzpicture}
				\centering
				\begin{axis}[legend pos=outer north east,legend to name=leg, width=8cm, height=6.75cm,legend entries={No Immunization, Random, Eigenvector Cent., Harmonic Cent., Betweenness Cent., Degree Cent., Comm. Bridge, Genetic Alg., Stochastic Prog.}, xtick=data, ylabel=Contact Infections, xlabel=Immunization Budget, cycle list name=methodsList]
					\addplot table [col sep=semicolon,x=Rate,y=NO IMMUNIZATION] {immunizationResults_dtu_eng_tech_averages.csv};
					
					\addplot table [col sep=semicolon, x=Rate,y=RANDOM] {immunizationResults_dtu_eng_tech_averages.csv};
					
					\addplot table [col sep=semicolon, x=Rate,y=EIGENVECTOR CENTRALITY] {immunizationResults_dtu_eng_tech_averages.csv};
					
					\addplot table [col sep=semicolon, x=Rate,y=HARMONIC CENTRALITY] {immunizationResults_dtu_eng_tech_averages.csv};
					
					\addplot table [col sep=semicolon, x=Rate,y=BETWEENNESS CENTRALITY] {immunizationResults_dtu_eng_tech_averages.csv};
					
					\addplot table [col sep=semicolon, x=Rate,y=DEGREE CENTRALITY ITERATIVE] {immunizationResults_dtu_eng_tech_averages.csv};
					
					\addplot table [col sep=semicolon, x=Rate,y=COMMUNITY BRIDGE FINDER] {immunizationResults_dtu_eng_tech_averages.csv};
					
					\addplot table [col sep=semicolon, x=Rate,y=GENETIC ALGORITHM] {immunizationResults_dtu_eng_tech_averages.csv};
					
					\addplot table [col sep=semicolon, x=Rate,y=STOCHASTIC PROGRAMMING] {immunizationResults_dtu_eng_tech_averages.csv};
				\end{axis}
			\end{tikzpicture}
			
			\caption{\revised{\DTUEngTech{}}}
		\end{subfigure}
		\begin{subfigure}{0.49\textwidth}
			\begin{tikzpicture}
				\centering
				\begin{axis}[legend to name=leg2, width=8cm, height=6.75cm, legend entries={No Immunization, Random, Eigenvector Cent., Harmonic Cent., Betweenness Cent., Degree Cent., Comm. Bridge, Genetic Alg., Stochastic Prog.}, xtick=data, ylabel=Contact Infections, xlabel=Immunization Budget,cycle list name=methodsList]
					\addplot table [col sep=semicolon, x=Rate,y=NO_IMMUNIZATION] {dtu_management_averages.csv};
					
					\addplot table [col sep=semicolon, x=Rate,y=RANDOM] {.//dtu_management_averages.csv};
					
					\addplot table [col sep=semicolon, x=Rate,y=EIGENVECTOR_CENTRALITY] {.//dtu_management_averages.csv};
					
					\addplot table [col sep=semicolon, x=Rate,y=HARMONIC_CENTRALITY] {.//dtu_management_averages.csv};
					
					\addplot table [col sep=semicolon, x=Rate,y=BETWEENNESS_CENTRALITY] {.//dtu_management_averages.csv};
					
					\addplot table [col sep=semicolon, x=Rate,y=DEGREE_CENTRALITY_ITERATIVE] {.//dtu_management_averages.csv};
					
					\addplot table [col sep=semicolon, x=Rate,y=COMMUNITY_BRIDGE_FINDER] {.//dtu_management_averages.csv};
					
					\addplot table [col sep=semicolon, x=Rate,y=GENETIC_ALGORITHM] {.//dtu_management_averages.csv};
					
					\addplot table [col sep=semicolon, x=Rate,y=STOCHASTIC_PROGRAMMING] {.//dtu_management_averages.csv};
				\end{axis}
			\end{tikzpicture}
			\caption{\DTUManagement{}}
		\end{subfigure}
		
		\vspace{0.3cm}
		
		\begin{subfigure}{0.5\textwidth}
			\begin{tikzpicture}
				\begin{axis}[legend to name=leg3, width=8cm, height=6.75cm, legend entries={No Immunization, Random, Eigenvector Cent., Harmonic Cent., Degree Cent., Comm. Bridge, Genetic Alg., Stochastic Prog.}, xtick=data, ylabel=Contact Infections, xlabel=Immunization Budget,cycle list name=methodsList]
					\addplot table [col sep=semicolon, x=Rate,y=NO IMMUNIZATION] {dtu_full_averages.csv};
					
					\addplot table [col sep=semicolon, x=Rate,y=RANDOM] {dtu_full_averages.csv};
					
					\addplot table [col sep=semicolon, x=Rate,y=EIGENVECTOR CENTRALITY] {dtu_full_averages.csv};
					
					\addplot table [col sep=semicolon, x=Rate,y=HARMONIC CENTRALITY] {dtu_full_averages.csv};
					
					\addplot table [col sep=semicolon, x=Rate,y=HARMONIC CENTRALITY] {dtu_full_averages.csv};
					
					\addplot table [col sep=semicolon, x=Rate,y=DEGREE CENTRALITY ITERATIVE] {dtu_full_averages.csv};
					
					\addplot table [col sep=semicolon, x=Rate,y=COMMUNITY BRIDGE FINDER] {dtu_full_averages.csv};
					
					\addplot table [col sep=semicolon, x=Rate,y=GENETIC ALGORITHM] {dtu_full_averages.csv};
					
					\addplot table [col sep=semicolon, x=Rate,y=STOCHASTIC PROGRAMMING] {dtu_full_averages.csv};
				\end{axis}
			\end{tikzpicture}
			
			\caption{Full DTU}
		\end{subfigure}
		\begin{subfigure}{0.49\textwidth}
			\centering
			\begin{tikzpicture}
				\ref{leg}
			\end{tikzpicture}
			\vspace{2cm}
			
			\caption{Legend}
		\end{subfigure}
	
		\caption{Results of the immunization methods for different immunization rates at a fixed contact infection probability $\contactSpreadProb = 0.25$.}
		\label{fig: performance over immunization budgets}
	\end{figure}

	\begin{figure}[htbp]
		\centering
		\begin{subfigure}{0.49\textwidth}
			\centering
			\begin{tikzpicture}
				\centering
				\begin{axis}[legend to name=leg, width=8cm, height=6.75cm,legend entries={No Immunization, Random, Eigenvector Cent., Harmonic Cent., Betweenness Cent., Degree Cent., Comm. Bridge, Genetic Alg., Stochastic Prog.}, xtick=data, ylabel=Contact Infections, xlabel=$\contactSpreadProb$,cycle list name=methodsList]
					\addplot table [col sep=semicolon, x=CI,y=NO_IMMUNIZATION] {dtu_eng_tech_ci.csv};
					
					\addplot table [col sep=semicolon, x=CI,y=RANDOM] {dtu_eng_tech_ci.csv};
					
					\addplot table [col sep=semicolon, x=CI,y=EIGENVECTOR_CENTRALITY] {dtu_eng_tech_ci.csv};
					
					\addplot table [col sep=semicolon, x=CI,y=HARMONIC_CENTRALITY] {dtu_eng_tech_ci.csv};
					
					\addplot table [col sep=semicolon, x=CI,y=BETWEENNESS_CENTRALITY] {dtu_eng_tech_ci.csv};
					
					\addplot table [col sep=semicolon, x=CI,y=DEGREE_CENTRALITY_ITERATIVE] {dtu_eng_tech_ci.csv};
					
					\addplot table [col sep=semicolon, x=CI,y=COMMUNITY_BRIDGE_FINDER] {dtu_eng_tech_ci.csv};
					
					\addplot table [col sep=semicolon, x=CI,y=GENETIC_ALGORITHM] {dtu_eng_tech_ci.csv};
					
					\addplot table [col sep=semicolon, x=CI,y=STOCHASTIC_PROGRAMMING] {dtu_eng_tech_ci.csv};
				\end{axis}
			\end{tikzpicture}
			
			\caption{\revised{\DTUEngTech{}}}
		\end{subfigure}
		\begin{subfigure}{0.49\textwidth}
			\centering
			\begin{tikzpicture}
				\centering
				\begin{axis}[legend to name=leg2, width=8cm, height=6.75cm, legend entries={No Immunization, Random, Eigenvector Cent., Harmonic Cent., Betweenness Cent., Degree Cent., Comm. Bridge, Genetic Alg., Stochastic Prog.}, xtick=data, ylabel=Contact Infections, xlabel=$\contactSpreadProb$,cycle list name=methodsList]
					\addplot table [col sep=semicolon, x=CI,y=NO_IMMUNIZATION] {dtu_management_ci.csv};
					
					\addplot table [col sep=semicolon, x=CI,y=RANDOM] {dtu_management_ci.csv};
					
					\addplot table [col sep=semicolon, x=CI,y=EIGENVECTOR_CENTRALITY] {dtu_management_ci.csv};
					
					\addplot table [col sep=semicolon, x=CI,y=HARMONIC_CENTRALITY] {dtu_management_ci.csv};
					
					\addplot table [col sep=semicolon, x=CI,y=BETWEENNESS_CENTRALITY] {dtu_management_ci.csv};
					
					\addplot table [col sep=semicolon, x=CI,y=DEGREE_CENTRALITY_ITERATIVE] {dtu_management_ci.csv};
					
					\addplot table [col sep=semicolon, x=CI,y=COMMUNITY_BRIDGE_FINDER] {dtu_management_ci.csv};
					
					\addplot table [col sep=semicolon, x=CI,y=GENETIC_ALGORITHM] {dtu_management_ci.csv};
					
					\addplot table [col sep=semicolon, x=CI,y=STOCHASTIC_PROGRAMMING] {./dtu_management_ci.csv};
				\end{axis}
			\end{tikzpicture}
			\caption{\DTUManagement{}}
		\end{subfigure}
		
		\vspace{0.3cm}
		
		\begin{subfigure}{0.49\textwidth}
			\centering
			\begin{tikzpicture}
				\begin{axis}[legend to name=leg3, width=8cm, height=6.75cm, legend entries={No Immunization, Random, Eigenvector Cent., Harmonic Cent., Betweenness Cent., Degree Cent., Comm. Bridge, Genetic Alg., Stochastic Prog.}, xtick=data, ylabel=Contact Infections, xlabel=$\contactSpreadProb$,cycle list name=methodsList]
					\addplot table [col sep=semicolon, x=CI,y=NO_IMMUNIZATION] {dtu_full_ci.csv};
					
					\addplot table [col sep=semicolon, x=CI,y=RANDOM] {dtu_full_ci.csv};
					
					\addplot table [col sep=semicolon, x=CI,y=EIGENVECTOR_CENTRALITY] {dtu_full_ci.csv};
					
					\addplot table [col sep=semicolon, x=CI,y=HARMONIC_CENTRALITY] {dtu_full_ci.csv};
					
					\addplot table [col sep=semicolon, x=CI,y=BETWEENNESS_CENTRALITY] {dtu_full_ci.csv};
					
					\addplot table [col sep=semicolon, x=CI,y=DEGREE_CENTRALITY_ITERATIVE] {dtu_full_ci.csv};
					
					\addplot table [col sep=semicolon, x=CI,y=COMMUNITY_BRIDGE_FINDER] {dtu_full_ci.csv};
					
					\addplot table [col sep=semicolon, x=CI,y=GENETIC_ALGORITHM] {dtu_full_ci.csv};
					
					\addplot table [col sep=semicolon, x=CI,y=STOCHASTIC_PROGRAMMING] {dtu_full_ci.csv};
				\end{axis}
			\end{tikzpicture}
			
			\caption{Full DTU}
		\end{subfigure}
		\begin{subfigure}{0.49\textwidth}
			\centering
			\ref{leg}
			\vspace{2cm}
			\caption{Legend}
		\end{subfigure}

		\caption{Results of the immunization methods for different contact infection probabilities at a fixed immunization rate of 20\%.}
		\label{fig: performance over CI}
	\end{figure}

	The immunization results in \autoref{fig: performance over immunization budgets} show that the stochastic programming heuristic performs well over the different contact graphs and immunization rates.
	This is especially the case for the \revised{\DTUEngTech{}} and \DTUFull{} graphs, for which the number of contact infections resulting from this method is the lowest among all methods for each of the immunization rates.
	The results of this method are more mixed for the \DTUManagement{} graph, but here the method is still among the best performing methods for each immunization rate.
	The genetic algorithm is also consistently among the best immunization methods.
	However, unlike the stochastic programming heuristic, it never obtains the best result of all immunization methods.
	
	These results on the performance of the stochastic programming heuristic and genetic algorithm are mostly confirmed by \autoref{fig: performance over CI} that considers  different contact infection probabilities.
	Again, the stochastic programming heuristic performs best on the \revised{\DTUEngTech{}} and \DTUFull{} graph, but more mixed on the \revised{\DTUManagement{}} graph.
	A potential explanation for this might be the very connected nature of the \DTUManagement{} graph, which might imply a better performance of the centrality-based measures and, specifically, the betweenness centrality method.
	It can again be seen that the genetic algorithm is consistently among the best methods but never exceeds all others.
	This result is somewhat surprising, especially as the genetic algorithm considers an initial population consisting of some solutions that are based on those of the considered centrality measures.
	
	When considering the results of the benchmark methods, it can be seen that these are also relatively stable over the different instances and parameter settings.
	Betweenness centrality particularly performs well over all instances and obtains the best result for many of the parameter settings of the \DTUManagement{} graph.
	The Degree centrality and the Community Bridge Finder method also perform well, where the latter shows particularly good performance on the smaller department contact graphs.
	Eigenvector centrality and Harmonic centrality perform less well and are often unable to outperform the random choice of nodes to be immunized.
	It should be noted here, though, that the best out of 10 random solutions is considered in the latter method.

	\subsubsection{Comparing the Immunized Nodes}
	To obtain insight into how the solutions from the different immunization methods differ, we use the Jaccard similarity measure.
	Given any two sets of immunized nodes $X, Y \subseteq V$, this measure can be computed as
	\begin{equation}
		\textcolor{black}{\textrm{Jaccard}(X,Y) = \frac{\vert X \cap Y \vert}{\vert X \cup Y \vert}.}
	\end{equation}
	The Jaccard similarity measure thus computes the fraction of common nodes in relation to the total number of unique nodes over both solutions.
	We have computed the Jaccard similarity measure between any two solutions obtained in the experiments in \autoref{fig: performance over CI} for the full contact graph.
	Note that these solutions of the immunization methods can differ over the contact infection probabilities, meaning that we compute the average over the similarity scores.
	The resulting scores are given in \autoref{tab: jaccard}.

	\begin{table}[htbp]
		\caption{Jaccard similarity score between the solutions of any two immunization methods, averaged over the three instances in \autoref{fig: performance over CI}.}
		\label{tab: jaccard}
		
		\resizebox{\textwidth}{!}{%
		\pgfplotstabletypeset[
		color cells={min=0,max=1},
		col sep=&,
		row sep=crcr,
		columns/Algorithm/.style={reset styles,string type},
		every head row/.style={
			before row=\toprule,
			after row=\midrule,
		},
		every last row/.style={
			after row=\bottomrule}
		]{
		Algorithm                     & Random       & Eigenvector & Harmonic & Betweenness & Degree & Comm. Bridge & GA & Stoch. \\
		Random                        & 1            & 0.1118933619            & 0.1108238527         & 0.1153313952            & 0.1160684004                  & 0.1065287053              & 0.1128573501       & 0.1149909007            \\
		Eigenvector       & 0.1118933619 & 1                       & 0.6530348687         & 0.3483146067            & 0.5232050774                  & 0.1749207465              & 0.181731503        & 0.1305223625            \\
		Harmonic          & 0.1108238527 & 0.6530348687            & 1                    & 0.4436090226            & 0.6006669446                  & 0.2021430884              & 0.1816623736       & 0.1342034592            \\
		Betweenness       & 0.1153313952 & 0.3483146067            & 0.4436090226         & 1                       & 0.4523449319                  & 0.2511319524              & 0.179129891        & 0.15178842              \\
		Degree & 0.1160684004 & 0.5232050774            & 0.6006669446         & 0.4523449319            & 1                             & 0.2148096524              & 0.1850786788       & 0.1606043604            \\
		Comm. Bridge     & 0.1065287053 & 0.1749207465            & 0.2021430884         & 0.2511319524            & 0.2148096524                  & 1                         & 0.1458137231       & 0.1307493081            \\
		GA            & 0.1128573501 & 0.181731503             & 0.1816623736         & 0.179129891             & 0.1850786788                  & 0.1458137231              & 1                  & 0.1312995057            \\
		Stoch.       & 0.1149909007 & 0.1305223625            & 0.1342034592         & 0.15178842              & 0.1606043604                  & 0.1307493081              & 0.1312995057       & 1                      \\
		}}
	\end{table}
	
	The table shows that the similarity scores overall are relatively low.
	The highest scores are obtained for solutions of the Harmonic centrality measure, for which about 2/3 of the immunized nodes \revised{coincide} with the Eigenvector centrality measure.
	Similarly, about 60\% and 45\% of the immunized nodes for the Harmonic centrality measure overlap with the Degree and Betweenness centrality measures, respectively.
	Unsurprisingly, the lowest similarity scores are obtained for the random selection of nodes, where, on average, about 11\% of the nodes are common with any other solution method.
	
	When looking at our newly proposed immunization methods, it can be seen that the GA solutions have the highest similarity score to the Eigenvector, Harmonic, Betweenness,  and Degree centrality solutions.
	This can likely be explained by some of these centrality measures being used to generate initial solutions within the GA. 
	However, even for these measures, on average, only about one in five nodes is in common with the solutions of the GA.
	The similarity scores for the stochastic programming heuristic are even a bit lower.
	Even compared to the best-performing centrality measures, which often obtain scores not very far away from those of the stochastic programming heuristic, no more than 16\% of the nodes is shared to these measures.
	Hence, these results suggest that there is a large number of relatively similar nodes in the DTU contact graph, which can be exchanged in solutions without very strongly impacting the immunization result.

	\subsection{\revised{Configuration and} Performance of the Stochastic Programming Heuristic and Genetic Algorithm}
	\label{sec: method analysis}
	\revised{Both the stochastic programming heuristic and GA come with parameters that need to be carefully selected to obtain the best possible performance.
	In this section, we look at how the choice of these parameters impacts the performance of these methods and further zoom in on the convergence of the GA.}
	
	\subsubsection{Selection of Number of Samples in Stochastic Programming Heuristic}
	\label{sec: SP samples}
	\revised{First}, we analyze the impact of the number of sampled infection forests $\sigma_p$ on the performance of the stochastic programming heuristic.
	\revised{In particular, we analyze the resulting number of contact infections, execution time of the method, and optimality gap in solving the MILP model for values of $\sigma_p$ between 100 and 1500.}
	The results of these experiments are given in \autoref{fig: stoch perf} for all three contact graphs.
	These results are based on a contact infection probability of $\contactSpreadProb = 0.35$ and immunization rate of 20\%.
	\revised{Moreover, the total time limit for the method has been extended to 5 hours, and the number of available processor cores to 12, to get a better image of the performance of the method for higher sample sizes.}
	
	\begin{figure}[htbp]
		\pgfplotstableread[col sep=semicolon]{
			Scenarios;Eng Tech inf;Eng Tech tim;Eng Tech gap;Management inf;Management tim;Management gap;Full inf;Full tim;Full gap
			100;32.65963287;25;0.145;40.82;30;0.2246;483.915;273;0.4794
			300;32.445;74;0.478;28.70995719;77;0.303;478.28;18156;26.1206
			500;30.115;3107;0.4959;26.02;1173;0.4948;450.08;18155;25.028
			700;30.865;18015;0.6656;24.105;2019;0.4929;431.625;18132;26.2295
			900;29.71;18016;1.5011;24.82;18019;0.6076;428.845;18163;26.0735
			1100;28.755;18017;2.4661;24.315;18018;0.6492;427.575;18160;26.0985
			1300;29.585;18009;3.2221;22.735;18019;1.3266;414.72;18155;26.4
			1500;30.595;18009;5.2372;22.28;18019;1.117;424.9;18163;27.3
		}{\stochData}

		\begin{subfigure}{0.35\textwidth}
			\centering
			\begin{tikzpicture}
				\begin{axis}[width=5.5cm,height=4.15cm,ylabel=Contact Inf.]
					\addplot table [col sep=semicolon, x=Scenarios,y=Eng Tech inf] \stochData;
				\end{axis}
		\end{tikzpicture}
		\end{subfigure}
		\begin{subfigure}{0.315\textwidth}
			\centering
			\begin{tikzpicture}
				\begin{axis}[width=5.5cm,height=4.15cm]
					\addplot table [col sep=semicolon, x=Scenarios,y=Management inf] \stochData;
				\end{axis}
			\end{tikzpicture}
		\end{subfigure}
		\begin{subfigure}{0.315\textwidth}
			\centering
			\begin{tikzpicture}
				\begin{axis}[width=5.5cm,height=4.15cm]
					\addplot table [col sep=semicolon, x=Scenarios,y=Full inf] \stochData;
				\end{axis}
			\end{tikzpicture}
		\end{subfigure}
		
		\vspace{0.2cm}
		
		\begin{subfigure}{0.35\textwidth}
			\begin{tikzpicture}
				\centering
				\begin{axis}[width=5.5cm,height=4.15cm,ylabel=Execut.\ Time (s)]
					\addplot table [col sep=semicolon, x=Scenarios,y=Eng Tech tim] \stochData;
				\end{axis}
			\end{tikzpicture}
		\end{subfigure}
		\begin{subfigure}{0.315\textwidth}
			\centering
			\begin{tikzpicture}
				\begin{axis}[width=5.5cm,height=4.15cm]
					\addplot table [col sep=semicolon, x=Scenarios,y=Management tim] \stochData;;
				\end{axis}
			\end{tikzpicture}
		\end{subfigure}
		\begin{subfigure}{0.315\textwidth}
			\centering
			\begin{tikzpicture}
				\begin{axis}[width=5.5cm,height=4.15cm]
					\addplot table [col sep=semicolon, x=Scenarios,y=Full tim] \stochData;
				\end{axis}
			\end{tikzpicture}
		\end{subfigure}
		
		\vspace{0.2cm}
		
		\begin{subfigure}{0.35\textwidth}
			\centering
			\begin{tikzpicture}
				\begin{axis}[width=5.5cm,height=4.15cm,ylabel=Optim.\ Gap (\%),xlabel=$\sigma_p$]
					\addplot table [col sep=semicolon, x=Scenarios, y expr=\thisrow{Eng Tech gap}] \stochData;
				\end{axis}
			\end{tikzpicture}
			
			\caption{\DTUEngTech{}}
		\end{subfigure}
		\begin{subfigure}{0.315\textwidth}
			\centering
			\begin{tikzpicture}
				\begin{axis}[width=5.5cm,height=4.15cm,xlabel=$\sigma_p$]
					\addplot table [col sep=semicolon, x=Scenarios, y expr=\thisrow{Management gap}] \stochData;;
				\end{axis}
			\end{tikzpicture}
			
			\caption{\DTUManagement{}}
		\end{subfigure}
		\begin{subfigure}{0.315\textwidth}
			\centering
			\begin{tikzpicture}
				\begin{axis}[width=5.5cm,height=4.15cm,xlabel=$\sigma_p$]
					\addplot table [col sep=semicolon, x=Scenarios, y expr=\thisrow{Full gap}] \stochData;
				\end{axis}
			\end{tikzpicture}
			
			\caption{Full DTU graph}
		\end{subfigure}
		
		\caption{Performance of the stochastic programming heuristic, in terms of the number of \revised{obtained} contact infections\revised{,} execution time\revised{, and optimality gap of the solver} for different values of $\sigma_p$.}
		\label{fig: stoch perf}
	\end{figure}
	
	The results in \autoref{fig: stoch perf} show that the performance of the stochastic programming \revised{heuristic generally improves} as the number of sampled infection forests increases.
	In particular, \revised{a strong decrease in the number of contact infections can be seen for all contact graphs when increasing the number of sampled infection forests from 100 to around 500 -- 700.
	Moreover, the lowest number of contact infections is obtained for each contact graph in the range of 1100 -- 1500 sampled infection forests.
	The performance for the highest number of sampled infection forests differs between the contact graph, though.
	For the \DTUEngTech{} graph, a small increase in the number of contact infections can be seen, while for the other graphs the number of infections seems to stabilize.
	}
	
	\revised{
		While the results show a decrease in the number of contact infections when increasing the number of sampled infection forests, one can also see that the execution time of the method grows quickly.	
		Especially the solving time of the MILP contributes to this, as the time needed for sampling only increases linearly in the number of infection forests.
		The rise in execution time} is especially clear for the full DTU contact graph, where the time limit of \revised{five} hours is already reached at \revised{300} sampled infection forests.
	\revised{One can also see a corresponding increase in the optimality gap, which goes up for all contact graphs and reaches up to about 25\% for the \DTUFull{} contact graph.
	This increasing optimality gap of the MILP solver likely contributes to the stabilization, or small increase, that can be seen in the number of contact infections at a higher number of sampled infection forests.}

%
%	While this leads to the method finding non-optimal IP solutions, often with significant optimality gaps \revised{of up to about 25\%}, this does not prevent the overall result from improving until a significantly larger number of infection forests is used.
%	Hence, these results show that a high number of samples is required to gain an adequate overview of the typical flow of infections and, thus, of the most important nodes to immunize.
%	
	\subsubsection{\revised{Selection of the Genetic Algorithm Parameters}}
	\label{sec: GA tuning}
	
	\begin{Revised}
		Second, we look at the selection of the evolutionary parameters used in our GA.
		In particular, we focus on the population size $\populationSize$,  mutation rate $\mutationRate$, tournament size $\tournamentSize$, and the number of individuals transferred between iterations through elitist selection $\epsilon$. 
		We perform our analysis by varying one parameter at a time from the base parameter setting in \autoref{tab:ga-parameterTable} that was determined through initial experimentation.
		The variations for each parameter, using a discretization with constant step size, are given in \autoref{tab: genetic tuning}.
		The performance of the genetic algorithm is evaluated for the \DTUEngTech{} graph, and we perform 5 independent runs of the genetic algorithm to limit the impact of the inherent stochastic variation that can be seen between subsequent runs of the algorithm for a given parameter setting.
		All experiments were conducted using a contact infection probability of $\contactSpreadProb{} = 0.35$ and immunization rate of 20\%.
		Moreover, to reduce the computational scope of these experiments, we used both a time-based stopping criterion of 90 minutes and an iteration-based limit of 1000 iterations for the GA, while at the same time again considering an increased number of 12 available processor cores.
		
		The results in \autoref{fig: ga parameter study} show that the GA's performance is indeed impacted by the choice of parameters, but that the performance of the GA is also relatively robust to the exact parameter setting chosen.
		The most clear pattern in terms of performance can be seen when varying the population size $\populationSize$ and mutation rate $\mutationRate$.
		For $\populationSize$, the best average performance of the GA can be seen at population sizes in the middle of the chosen range (75 -- 125) with a slight increase in the number of contact infections at the highest population size of 150.
		For the mutation rate $\rho$, low but non-zero values (0.025 - 0.075) show the best results, although there is an outlier to the pattern for $\rho = 0.15$.
		For the elitist selection size $\epsilon$, there seems to be evidence that elitist selection indeed aids performance, as the worst performance occurs when using no elitist selection ($\epsilon = 0$).
		The best results are obtained for $\epsilon = 4$, although larger values between 8 -- 10 also show good performance.
		For the tournament size, the pattern is less clear and largely overshadowed by random variation between the iteration runs.
	\end{Revised}
		
	\begin{table}[htbp]
		\caption{\revised{Overview of the considered GA parameter values.}}
			
		\centering
			
		\begin{Revised}
			\begin{tabular}{lrrrr}
				\toprule
				Parameter & Base value & \multicolumn{3}{c}{Parameter Variations} \\
				\cmidrule(lr){3-5}
				& & Lower value & Upper value & Step size \\
				\midrule
				$\populationSize$ & 100 & 25 & 150 & 25 \\
				$\mutationRate$ & 0.05 & 0 & 0.15 & 0.025 \\
				$\tournamentSize$ & 4 & 2 & 10 & 2 \\
				$\epsilon$ & 5 & 0 & 10 & 2 \\
				\bottomrule
			\end{tabular}
		\end{Revised}

		\label{tab: genetic tuning}
	\end{table}
	
	\begin{figure}[htbp]
		\begin{subfigure}{0.49\textwidth}
			\centering
			\begin{tikzpicture}
				\begin{axis}[height=4.75cm,width=7cm,ymin=56,ymax=64,ylabel=Cont. Inf.]
					
					\foreach \i in {1,...,5}{
						\addplot[blue,only marks,opacity=0.5] table [col sep=semicolon,x=ps,y=\i] {genTuningPopulationRepetition.csv};
					
					}
					
					\addplot[line width=0.6pt,black,densely dashed,mark=star,mark options={solid}] table [col sep=semicolon,x=ps,y=Average] {genTuningPopulationRepetition.csv};
				\end{axis}
			\end{tikzpicture}
			
			\caption{\revised{Population Size $N$}}
		\end{subfigure}
		\begin{subfigure}{0.49\textwidth}
			\centering
			\begin{tikzpicture}
				\pgfplotstableread[col sep=semicolon]{
					mr;1;2;3;4;5;Average
					0;60.465;59.9;61.715;59.25;60.825;60.431
					0.025;59.595;61.43;57.785;61.3;59.285;59.879
					0.05;58.665;60.935;61.375;58.235;59.895;59.821
					0.075;60.235;58.875;60.335;61.785;61.635;60.573
					0.1;59.795;59.7;61.91;63.645;59.055;60.821
					0.125;62.015;62.995;60.255;62.57;62.4;62.047
					0.15;61.39;59.555;60.245;58.605;61.55;60.269
				}{\genTuningMutation}

				\begin{axis}[height=4.75cm,width=7cm,ymin=56,ymax=64,xticklabel style={/pgf/number format/fixed},ylabel=Cont. Inf.]
					
					\foreach \i in {1,...,5}{
						\addplot[blue,only marks,opacity=0.5] table [col sep=semicolon,x=mr,y=\i] \genTuningMutation;
						
					}
					
					\addplot[line width=0.6pt,black,densely dashed,mark=star,mark options={solid}] table [col sep=semicolon,x=mr,y=Average] \genTuningMutation;
				\end{axis}
			\end{tikzpicture}
		
			\caption{\revised{Mutation Rate $\rho$}}
		\end{subfigure}

		\begin{subfigure}{0.49\textwidth}
			\pgfplotstableread[col sep=semicolon]{
				ts;1;2;3;4;5;Average
				2;59.085;60.815;58.975;59.91;60.22;59.801
				4;60.7;61.09;59.58;62.425;58.895;60.538
				6;59.405;61.215;58.565;58.605;60.045;59.567
				8;59.065;60.575;60.04;60.24;60.175;60.019
				10;61.055;59.55;59.085;59.525;57.985;59.44
			}{\genTuningTournament}

			\centering
			\begin{tikzpicture}
				\begin{axis}[height=4.75cm,width=7cm,ymin=56,ymax=64,ylabel=Cont. Inf.]
					
					\foreach \i in {1,...,5}{
						\addplot[blue,only marks,opacity=0.5] table [col sep=semicolon,x=ts,y=\i] \genTuningTournament;
						
					}
					
					\addplot[line width=0.6pt,black,densely dashed,mark=star,mark options={solid}] table [col sep=semicolon,x=ts,y=Average,row sep=crcr] \genTuningTournament;
				\end{axis}
			\end{tikzpicture}
			
			\caption{\revised{Tournament Size $\tournamentSize$}}
		\end{subfigure}
		\begin{subfigure}{0.49\textwidth}
			\centering
			\begin{tikzpicture}
				\begin{axis}[height=4.75cm,width=7cm,ymin=56,ymax=64,ylabel=Cont. Inf.]
					
					\foreach \i in {1,...,5}{
						\addplot[blue,only marks,opacity=0.5] table [col sep=semicolon,x=es,y=\i] {genTuningElitistRepetition.csv};
						
					}
					
					\addplot[line width=0.6pt,black,densely dashed,mark=star,mark options={solid}] table [col sep=semicolon,x=es,y=Average,] {genTuningElitistRepetition.csv};
				\end{axis}
			\end{tikzpicture}
			
			\caption{\revised{Elitist Size $\epsilon$}}
		\end{subfigure}
		\caption{\revised{Performance of the GA, in terms of the number of obtained contact infections, for different values of the evolutionary parameters. The average performance is given by the dashed line, while the blue dots indicate the individual results of the 5 replications per parameter setting.}}
		\label{fig: ga parameter study}
	\end{figure}
	
	\subsubsection{Convergence of Genetic Algorithm}
	\revised{We also look at the convergence of the GA for the original parameter setting in \autoref{tab:ga-parameterTable}, which is shown in \autoref{fig: convergence genetic algorithm} for the experiments for the \DTUFull{} graph of \autoref{fig: performance over CI}}.
	In these plots, both the average score over all solutions in the population and the score of the best solution so far are given at each iteration.
	Note that the former is based on the evaluation in the small simulations, while the latter is based on the evaluation of the large simulation.
	Hence, the score of the best solution can be higher than the average score of the solutions.
	
		\begin{figure}[htbp]
		
		\centering
		
		\begin{subfigure}{0.35\textwidth}
			\begin{tikzpicture}
				\begin{axis}[xmin=2,width=5.5cm,height=4.25cm,ylabel=Contact Inf.,xlabel=Iteration]
					
					\addplot[line width=0.6pt,blue,densely dashed] table [col sep=comma,x=iteration,y=average,mark=none] {GA_performance_dtu_full_0.15_0.2_0.5_0.4.csv};
					
					\addplot[line width=0.6pt,red] table [col sep=comma,x=iteration,y=best,mark=none] {GA_performance_dtu_full_0.15_0.2_0.5_0.4.csv};
				\end{axis}
			\end{tikzpicture}
			
			\caption{$\contactSpreadProb = 0.15$}
		\end{subfigure}
		\begin{subfigure}{0.315\textwidth}
			\begin{tikzpicture}
				\begin{axis}[width=5.5cm,height=4.25cm,xmin=6,xlabel=Iteration]
					\addplot[line width=0.6pt,blue,densely dashed] table [col sep=comma,x=iteration,y=average,mark=none] {GA_performance_dtu_full_0.25_0.2_0.5_0.4.csv};
					
					\addplot[line width=0.6pt,red] table [col sep=comma,x=iteration,y=best,mark=none] {GA_performance_dtu_full_0.25_0.2_0.5_0.4.csv};
				\end{axis}
			\end{tikzpicture}
			
			\caption{$\contactSpreadProb = 0.25$}
		\end{subfigure}
		\begin{subfigure}{0.315\textwidth}
			\begin{tikzpicture}
				\begin{axis}[width=5.5cm,height=4.25cm,xmin=2,mark=none,xlabel=Iteration]
					\addplot[line width=0.6pt,blue,densely dashed] table [dashed,col sep=comma,x=iteration,y=average,mark=none] {GA_performance_dtu_full_0.35_0.2_0.5_0.4.csv};
					
					\addplot[line width=0.6pt,red] table [col sep=comma,x=iteration,y=best,mark=none] {GA_performance_dtu_full_0.35_0.2_0.5_0.4.csv};
				\end{axis}
			\end{tikzpicture}
			
			\caption{$\contactSpreadProb = 0.35$}
		\end{subfigure}

		\caption{Evolution of the contact infections over the iterations of the genetic algorithm, showing both the average contact infections for each generation (blue dashed line) as evaluated in the small simulations and the current best solution (red solid line) as evaluated in the large simulation.}
		\label{fig: convergence genetic algorithm}
	\end{figure}
	
	When looking at the progression of the genetic algorithm over the iterations in \autoref{fig: convergence genetic algorithm}, it can be seen that the best score clearly lags behind the average score.
	Moreover, the improvement in the number of infections of the best solution is significantly less than the improvement in the average score.
	Both can likely be explained by the larger number of simulation runs used in evaluating the best solution, which makes it harder to find improving solutions.
	When looking at the progression over the iterations, it can also be seen that the improvement clearly levels off for the average score but is not fully flat yet after the chosen computation time.
	Similarly, the frequency of finding a better solution decreases but some improving solutions are still found in the later iterations of the genetic algorithm.
	Hence, it can be seen that full convergence cannot be obtained within the set solution time of three hours and the clearly limited number of iterations that can be executed in this time.
	The limited number of iterations is thus likely a factor that strongly contributes to the mixed performance we see for the genetic algorithm, even though the number of iterations that can be performed increases to be in the hundreds for the smaller department graphs.

	\begin{Revised}
			\subsection{Impact of Disease Parameters}
			\label{sec: sensitivity}
			In this section, we zoom in on the impact of the disease parameters on the performance and solutions of the proposed immunization methods.
			Here, we investigate the effect of an activity-dependent infection probability and the robustness of the obtained solutions to changes in the disease parameters. 
			
			\subsubsection{Varying Contact Infection Probability per Activity}
			\label{sec: varying contact rate}
			So far, we have assumed that the contact infection probability $\contactSpreadProb$ is the same for all activities, i.e., for all hyperarcs $h \in H$.
			This reflects that it has been difficult in many environments to reliably estimate how the nature of an activity and its physical environment impact the spreading rate.
			However, in some settings, decision makers might have well-founded estimates of the infections per activity, e.g., when detailed contract tracing has been performed for repeating activities.
			Therefore, we also investigate how the performance of the methods and their solutions are affected by a contact infection probability $\contactSpreadProb^h$ that depends on the activity $h \in H$.
			In these experiments, we choose  $\contactSpreadProb^h$ for each activity $h \in H$ uniformly at random from the range introduced in \autoref{table:seir-model-param}, i.e., between 0.15 and 0.35.
			Moreover, we then determine if a person moves from the Susceptible to the Exposed state according to the probability $\contactSpreadProb^h$ of the activity $h \in H$ during which the contact with an infected person took place.
			The results of our experiments, in which we evaluate the performance of the methods for the same immunization rates as in \autoref{fig: performance over immunization budgets}, are shown in \autoref{fig: varying contact spread}.
			
			\begin{figure}[htbp]
				\begin{subfigure}{0.49\textwidth}
					\begin{tikzpicture}
						\centering
						\begin{axis}[legend pos=outer north east,legend to name=leg, width=8cm, height=6.25cm,legend entries={No Immunization, Random, Eigenvector Cent., Harmonic Cent., Betweenness Cent., Degree Cent., Comm. Bridge, Genetic Alg., Stochastic Prog.}, xtick=data, ylabel=Contact Infections, xlabel=Immunization Budget, cycle list name=methodsList]
							\addplot table [col sep=semicolon,x=Rate,y=NO_IMMUNIZATION] {immunizationResults_dtu_eng_tech_averages_varying_rates.csv};
							
							\addplot table [col sep=semicolon, x=Rate,y=RANDOM] {immunizationResults_dtu_eng_tech_averages_varying_rates.csv};
							
							\addplot table [col sep=semicolon, x=Rate,y=EIGENVECTOR_CENTRALITY] {immunizationResults_dtu_eng_tech_averages_varying_rates.csv};
							
							\addplot table [col sep=semicolon, x=Rate,y=HARMONIC_CENTRALITY] {immunizationResults_dtu_eng_tech_averages_varying_rates.csv};
							
							\addplot table [col sep=semicolon, x=Rate,y=BETWEENNESS_CENTRALITY] {immunizationResults_dtu_eng_tech_averages_varying_rates.csv};
							
							\addplot table [col sep=semicolon, x=Rate,y=DEGREE_CENTRALITY_ITERATIVE] {immunizationResults_dtu_eng_tech_averages_varying_rates.csv};
							
							\addplot table [col sep=semicolon, x=Rate,y=COMMUNITY_BRIDGE_FINDER] {immunizationResults_dtu_eng_tech_averages_varying_rates.csv};
							
							\addplot table [col sep=semicolon, x=Rate,y=GENETIC_ALGORITHM] {immunizationResults_dtu_eng_tech_averages_varying_rates.csv};
							
							\addplot table [col sep=semicolon, x=Rate,y=STOCHASTIC_PROGRAMMING] {immunizationResults_dtu_eng_tech_averages_varying_rates.csv};
						\end{axis}
					\end{tikzpicture}
					
					\caption{\revised{\DTUEngTech{}}}
				\end{subfigure}
				\begin{subfigure}{0.49\textwidth}
					\begin{tikzpicture}
						\centering
						\begin{axis}[legend to name=leg2, width=8cm, height=6.25cm, legend entries={No Immunization, Random, Eigenvector Cent., Harmonic Cent., Betweenness Cent., Degree Cent., Comm. Bridge, Genetic Alg., Stochastic Prog.}, xtick=data, ylabel=Contact Infections, xlabel=Immunization Budget,cycle list name=methodsList]
							\addplot table [col sep=semicolon, x=Rate,y=NO_IMMUNIZATION] {./dtu_management_averages_varying_rates.csv};
							
							\addplot table [col sep=semicolon, x=Rate,y=RANDOM] {./dtu_management_averages_varying_rates.csv};
							
							\addplot table [col sep=semicolon, x=Rate,y=EIGENVECTOR_CENTRALITY] {./dtu_management_averages_varying_rates.csv};
							
							\addplot table [col sep=semicolon, x=Rate,y=HARMONIC_CENTRALITY] {./dtu_management_averages_varying_rates.csv};
							
							\addplot table [col sep=semicolon, x=Rate,y=BETWEENNESS_CENTRALITY] {./dtu_management_averages_varying_rates.csv};
							
							\addplot table [col sep=semicolon, x=Rate,y=DEGREE_CENTRALITY_ITERATIVE] {./dtu_management_averages_varying_rates.csv};
							
							\addplot table [col sep=semicolon, x=Rate,y=COMMUNITY_BRIDGE_FINDER] {./dtu_management_averages_varying_rates.csv};
							
							\addplot table [col sep=semicolon, x=Rate,y=GENETIC_ALGORITHM] {./dtu_management_averages_varying_rates.csv};
							
							\addplot table [col sep=semicolon, x=Rate,y=STOCHASTIC_PROGRAMMING] {./dtu_management_averages_varying_rates.csv};
						\end{axis}
					\end{tikzpicture}
					\caption{\revised{\DTUManagement{}}}
				\end{subfigure}
				
				\vspace{0.3cm}
				
				\begin{subfigure}{0.5\textwidth}
					\begin{tikzpicture}
						\begin{axis}[legend to name=leg3, width=8cm, height=6.25cm, legend entries={No Immunization, Random, Eigenvector Cent., Harmonic Cent., Degree Cent., Comm. Bridge, Genetic Alg., Stochastic Prog.}, xtick=data, ylabel=Contact Infections, xlabel=Immunization Budget,cycle list name=methodsList]
							\addplot table [col sep=comma, x=Rate,y=NO_IMMUNIZATION] {dtu_full_averages_varying_rates.csv};
							
							\addplot table [col sep=comma, x=Rate,y=RANDOM] {dtu_full_averages_varying_rates.csv};
							
							\addplot table [col sep=comma, x=Rate,y=EIGENVECTOR_CENTRALITY] {dtu_full_averages_varying_rates.csv};
							
							\addplot table [col sep=comma, x=Rate,y=HARMONIC_CENTRALITY] {dtu_full_averages_varying_rates.csv};
							
							\addplot table [col sep=comma, x=Rate,y=HARMONIC_CENTRALITY] {dtu_full_averages_varying_rates.csv};
							
							\addplot table [col sep=comma, x=Rate,y=DEGREE_CENTRALITY_ITERATIVE] {dtu_full_averages_varying_rates.csv};
							
							\addplot table [col sep=comma, x=Rate,y=COMMUNITY_BRIDGE_FINDER] {dtu_full_averages_varying_rates.csv};
							
							\addplot table [col sep=comma, x=Rate,y=GENETIC_ALGORITHM] {dtu_full_averages_varying_rates.csv};
							
							\addplot table [col sep=comma, x=Rate,y=STOCHASTIC_PROGRAMMING] {dtu_full_averages_varying_rates.csv};
						\end{axis}
					\end{tikzpicture}
					
					\caption{\revised{Full DTU}}
				\end{subfigure}
				\begin{subfigure}{0.49\textwidth}
					\centering
					\begin{tikzpicture}
						\ref{leg}
					\end{tikzpicture}
					\vspace{2cm}
					
					\caption{\revised{Legend}}
				\end{subfigure}
				
				\caption{\revised{Results of the immunization methods for different immunization percentages with a varying contact infection probability per activity.}}
				\label{fig: varying contact spread}
			\end{figure}
			
			\autoref{fig: varying contact spread} shows that the comparison between the immunization methods is not affected in a major way when considering a varying contact infection probability.
			In particular, the stochastic programming heuristic is still able to obtain the best results for the \DTUEngTech{} and \DTUFull{} graphs, with more mixed results for the \DTUManagement{} graph.
			Moreover, it can be seen that the GA is again among the best-performing immunization methods.
			The GA performs slightly better, though, in relation to the other centrality measures than in our earlier experiments, where the GA now consistently gives the second-best results for the \DTUFull{} graph.
			This is likely an effect of the GA being able to leverage information about the infection rate of different activities through the simulation evaluations, which it runs during the search, while the solutions of the benchmark methods are not tuned to these probabilities.
			In line with the above, it can also be seen that the ranking of benchmark methods is largely in line with that seen with a fixed contact infection probability.
			
			When looking at the impact of the varying contact infection probability on the disease spread and the effect of immunization, we see that the results differ per instance.
			For the \DTUManagement{} graph, a small increase can be seen in the number of contact infections both without and with immunization, while no strong effect can be seen for the other graphs.
			Overall, the impact of immunization seems not to be strongly affected either.
			
			\subsubsection{Robustness of Solutions to Disease Parameters}
			\label{sec: robustness}
			Next, we look at how robust the solutions obtained by the immunization methods are to changes to the disease parameters.
			Our aim is, particularly, to evaluate the extent to which the performance that we saw for the solutions of our proposed immunization methods generalizes to disease parameters that differ from the ones for which they were obtained. 
			In our analysis, we consider the solutions, i.e., sets of immunized nodes, found by our stochastic programming heuristic and GA for the \DTUEngTech{} graph with a contact infection probability of $\beta_{\mathit{con}} = 0.25$ and immunization rate of 20\%.
            We re-evaluate the disease spread for these solutions through simulation for a large number of configurations of the disease parameters, where we vary the contact infection probability $\contactSpreadProb$, spontaneous infection probability $\spontaneousSpreadProb$, quarantine probability $\selfQuarantineProb$, and quarantine probability in case of close contact $\contactQuarProb$.
            We then perform a comparison to the number of infections obtained by the iterative version of the Degree centrality (\ref{meth: degree}) benchmark method, which is evaluated for the same disease parameter configurations.
            
            The results of our experiments are given in \autoref{fig: robustness stochastic} and \autoref{fig: robustness GA}.
            In these figures, each column and row corresponds to a single value of $\contactSpreadProb$ and $\spontaneousSpreadProb$, respectively.
            Moreover, each individual sub-plot, i.e., 9 by 9 heat map, gives the results for the different values of $\contactQuarProb$ and $\selfQuarantineProb$.
            Squares in blue show a lower number of infections of the proposed methods, while a square in red indicates a lower number of infections obtained by the solution of the iterative Degree Centrality benchmark.
			Note that the disease configuration for which these solutions were obtained can be found exactly in the center of the middle heatplot, i.e., the sub-plot for $\contactSpreadProb = 0.25$ and $\spontaneousSpreadProb = 0.0003$.
			Similar plots, but then for a comparison to the Random benchmark method, can be found in \autoref{sec: extra robustness figures}.
            
            \begin{figure}[htbp]
            	\centering
            	
            	\includegraphics[scale=0.65]{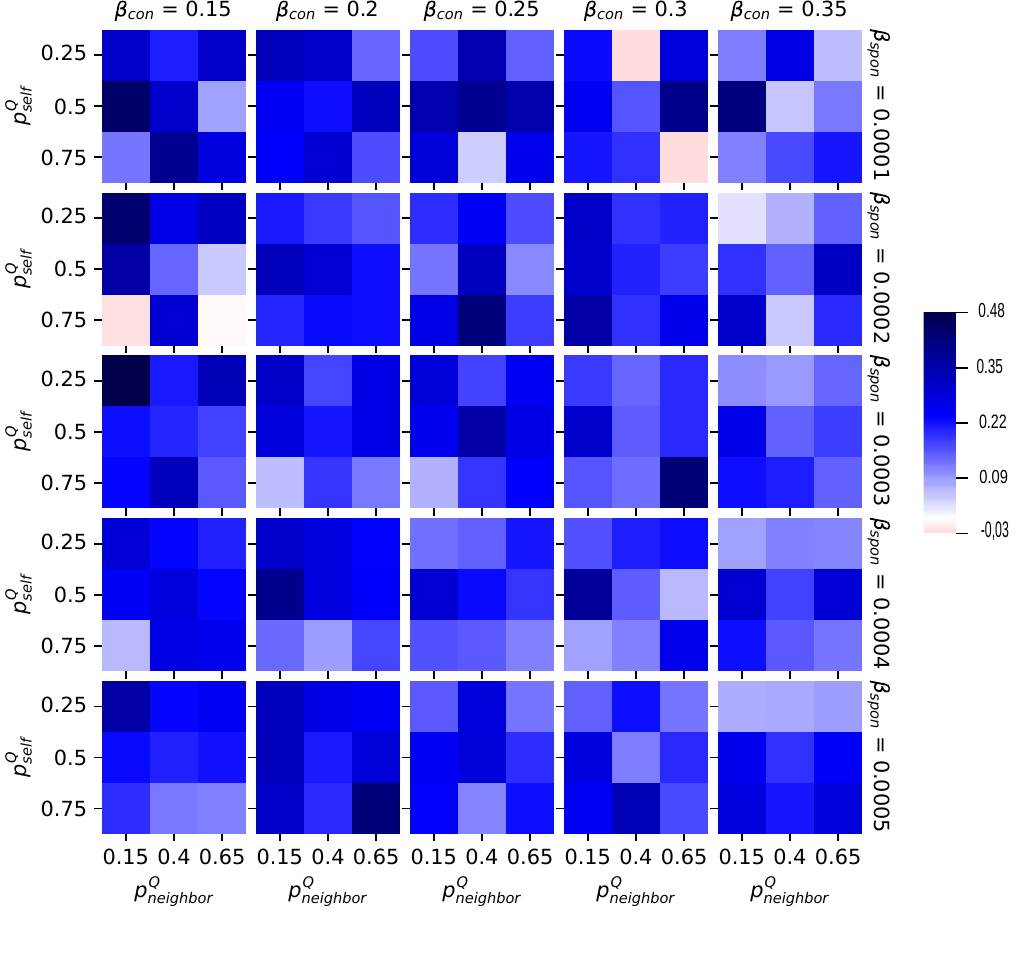}
            	
            	\vspace{-0.25cm}
            
            	\caption{\revised{Four-dimensional plot of the relative difference in the number of contact infections between the solutions of the stochastic programming heuristic and iterative Degree Centrality benchmark method for different disease parameter configurations.}}
            	\label{fig: robustness stochastic}
            \end{figure}
        
	        \begin{figure}[htbp]
	        	\centering
	        	\includegraphics[scale=0.65]{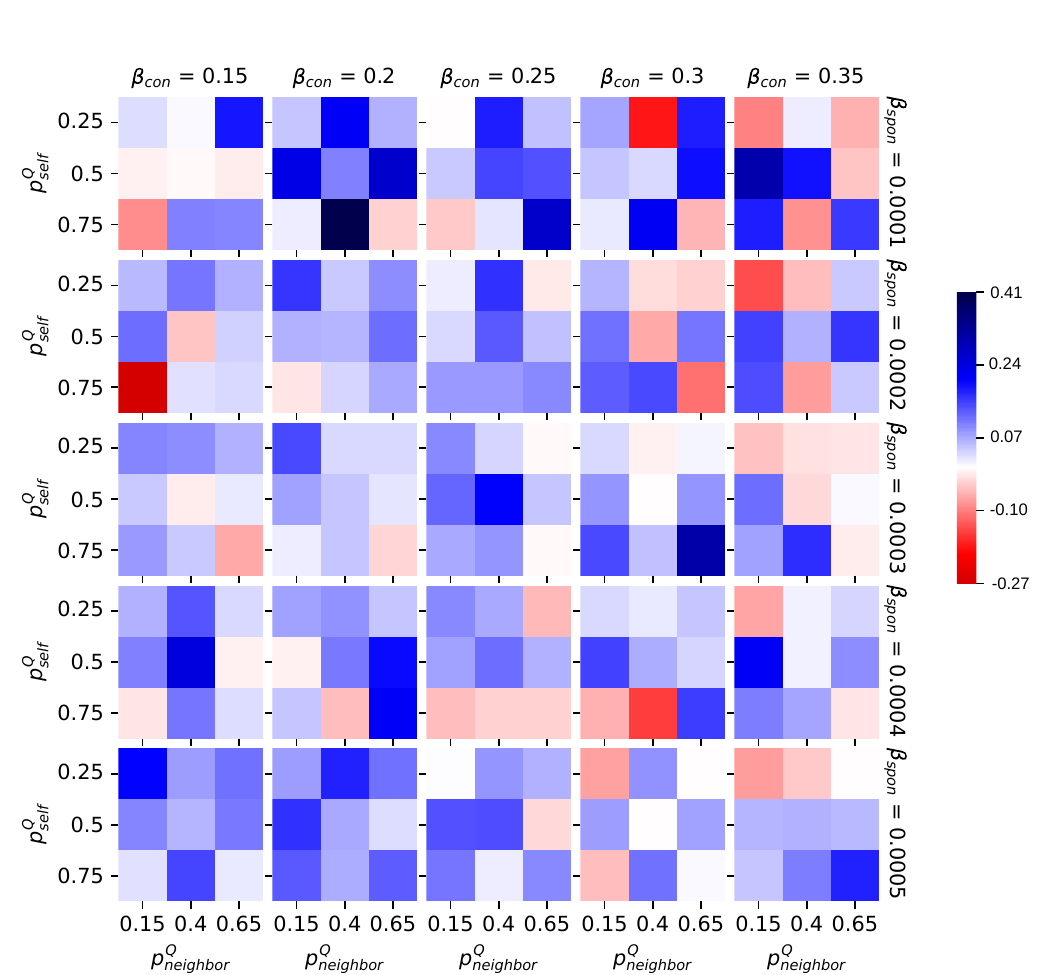}
	        	
	        	\caption{\revised{Four-dimensional plot of the relative difference in the number of contact infections between the solutions of the GA and iterative Degree Centrality benchmark method for different disease parameter configurations.}}
	        	\label{fig: robustness GA}
	        \end{figure}
        	
       		Overall, the results show that the improved performance of the stochastic programming heuristic, and to a lesser extent the GA, also generalizes to other disease parameter configurations.
       		As could be expected based on the stochastic programming solution giving the best performance for the original disease parameter configuration for which these solutions were obtained, the stochastic programming heuristic consistently shows the largest relative reduction compared to the iterative degree centrality benchmark.
       		In particular, it can be seen that there are only 4 parameter settings for which the stochastic programming heuristic solution does not give a relatively clear lower number of infections.
       		The results for the GA are more mixed, but the GA solution still leads to a lower number of contact infections than the iterative Degree centrality benchmark and the number of cases with a large reduction in infections for the GA clearly outnumbers the cases where the iterative Degree centrality benchmark leads to a large reduction.
       		For both methods, it can be seen that the relative improvements are generally lower for the highest value of the contact infection probability $\contactSpreadProb$, while larger relative reductions can more often be found at the lowest values of $\contactSpreadProb$.
       		Moreover, as one would expect, relative improvements are often realized when the parameters are close to the ones for which these solutions were obtained.
	\end{Revised}

	\subsection{Comparison to Minimizing Distinct Contacts}
	Lastly, we zoom in on the effect of immunization instead of focusing on the performance of the individual immunization methods.
	Here, we compare the effect of immunization with the effect of limiting group sizes, in which students are distributed over multiple smaller groups for large courses in such a way that the number of distinct contacts between students is minimized.
	We use the algorithm proposed in  \textcite{bagger2022reducing} to limit the group size for the complete DTU dataset to at most 50 and 30 students, resulting in two new contact graphs.
	Moreover, we evaluate the effect of combining the measures, where immunization is applied after first assigning students to these smaller course groups such to minimize the number of distinct contacts.	
	The corresponding disease spread for these different approaches is shown in \autoref{fig: immunization vs distinct contacts}, where the immunization results were obtained by using the stochastic programming heuristic and an immunization rate of 10\%, 20\%, or 30\%. 
	
	\begin{figure}[htbp]
		\centering
		\begin{tikzpicture}
			\begin{axis}[ytick={1,2,3,4,5,6,7,8,9,10,11,12}, yticklabels={Original Graph, Group Size 50, Group Size 30, 10\% Immunized, 10\% Immunized + Gr.\ Size 50, 10\% Immunized + Gr.\ Size 30, 20\% Immunized, 20\% Immunized + Gr.\ Size 50, 20\% Immunized + Gr.\ Size 30, 30\% Immunized, 30\% Immunized + Gr.\ Size 50, 30\% Immunized + Gr.\ Size 30},xlabel={Contact infections},width=10cm,height=8cm]
				\addplot+ [boxplot] table[col sep=comma, y=NO_IMMUNIZATION] {immunizationResults_dtu_full_0.25_0.2.csv};
				
				\addplot+ [boxplot] table[col sep=comma, y=NO_IMMUNIZATION] {immunizationResults_dtu_50_0.25_0.2.csv};
				
				\addplot+ [boxplot] table[col sep=comma, y=NO_IMMUNIZATION] {immunizationResults_dtu_30_0.25_0.2.csv};
				
				\addplot+ [boxplot] table[col sep=comma, y=STOCHASTIC_PROGRAMMING] {immunizationResults_dtu_full_0.25_0.1.csv};
				
				\addplot+ [boxplot] table[col sep=comma, y=STOCHASTIC_PROGRAMMING] {immunizationResults_dtu_50_0.25_0.1.csv};
				
				\addplot+ [boxplot] table[col sep=comma, y=STOCHASTIC_PROGRAMMING] {immunizationResults_dtu_30_0.25_0.1.csv};
				
				\addplot+ [boxplot] table[col sep=comma, y=STOCHASTIC_PROGRAMMING] {immunizationResults_dtu_full_0.25_0.2.csv};
				
				\addplot+ [boxplot] table[col sep=comma, y=STOCHASTIC_PROGRAMMING] {immunizationResults_dtu_50_0.25_0.2.csv};
				
				\addplot+ [boxplot] table[col sep=comma, y=STOCHASTIC_PROGRAMMING] {immunizationResults_dtu_30_0.25_0.2.csv};
				
				\addplot+ [boxplot] table[col sep=comma, y=STOCHASTIC_PROGRAMMING] {immunizationResults_dtu_full_0.25_0.3.csv};
				
				\addplot+ [boxplot] table[col sep=comma, y=STOCHASTIC_PROGRAMMING] {immunizationResults_dtu_50_0.25_0.3.csv};
				
				\addplot+ [boxplot] table[col sep=comma, y=STOCHASTIC_PROGRAMMING] {immunizationResults_dtu_30_0.25_0.3.csv};
			\end{axis}
		\end{tikzpicture}
		
		\caption{Comparison between network immunization and minimizing the number of distinct contacts. The number of contact infections obtained through simulation is given as a boxplot, where the centre line in each box gives the median number of infections.}
		\label{fig: immunization vs distinct contacts}
	\end{figure}
	
	The results in \autoref{fig: immunization vs distinct contacts} show that both the minimization of distinct contacts and network immunization clearly reduce the median number of contact infections.
	Reducing the group size to 50 or 30 students leads to a reduction of contact infections from about 550 infections in the original graph to about 475 and 450 infections, respectively.
	Immunization has an even larger effect, since the number of contact infections reduces to about 375, 275, and 225 for the different immunization rates, respectively.
	It can be concluded that immunization has the largest effect for the chosen parameters, as immunizing just 10\% of the students leads to a lower median number of infections than reducing the group size to a maximum of 30 students.	
	Moreover, one can see that the number of contact infections decreases sharply at higher immunization rates.
	
	The results in \autoref{fig: immunization vs distinct contacts} additionally show that the combination of minimizing distinct contacts and immunization can lead to a further reduction of infections. 
	A clear jump in the median number of infections can especially be seen when moving from immunization only to combining immunization and a maximum group size of 50 students.
	In comparison, the jump is smaller when the group size is further reduced to 30 students.
	This is in line with the results we saw for reducing group sizes only, where the jump from the original graph to a maximum of 50 students is also larger than the jump of moving from 50 to 30 students in a group.
	It should be noted that \revised{all pair-wise comparisons between the different approaches shown in \autoref{fig: immunization vs distinct contacts} are also statistically significant at the 1\% level (see \autoref{sec: stat distinct contacts}), even though significant variation exists between the experiments in a run.}
	
	\section{Conclusion \revised{and Discussion}}
	\label{sec: conclusion}
	
	In this paper, we looked at the \problemName{} in the context of an epidemic disease and \revised{a setting with scheduled activities}.
	This problem focuses on choosing individuals to immunize, given a maximum immunization rate, such that the spread of the disease through the population is minimized.
	Compared to the existing literature on this problem, we consider a richer epidemiological setting that includes the quarantining of infected individuals and their close contacts, and a limited willingness to test and quarantine.
	As a result, a simulation approach is used to evaluate the effect of an immunization strategy, where we focus on the number of infections that follow from contacts in the population.
	
	We proposed two simulation-optimization approaches for the \problemName{}: a stochastic programming heuristic and a genetic algorithm.
	The stochastic programming heuristic is based on sample average approximation, where we sample infection forests through simulation \revised{and afterwards} solve an optimization problem to choose the immunized nodes so that the number of infections that would occur in these infection forests is minimized.
	In the genetic algorithm, we use the results from existing centrality measures to choose an initial population and combine simulation runs of small and large size to balance the time needed to find promising solutions with the uncertainty that results from simulation.
	Moreover, we parallelized the simulation runs to increase the number of iterations that can be run.

	We applied the proposed algorithms to a contact graph based on students' course assignments for a major university in Denmark.
	This contact graph shows both small-world properties and a community structure.
	Our results show that our proposed methods are competitive with the best centrality measures and that the stochastic programming heuristic is able to outperform these immunization measures for a considerable number of these instances.
	We \revised{have} also compared the corresponding solutions of the immunization methods, \revised{showing that especially} the solutions of the stochastic programming method \revised{tend to differ from the centrality-based methods, and shown that the solutions are relatively robust to variations in the disease parameters.}  
	Furthermore, we looked at the effectiveness of immunization by comparing it to the strategy proposed in \textcite{bagger2022reducing} to minimize the number of distinct contacts when assigning students to course groups.
	Our experiments show that immunization can lead to a relatively large reduction in infections and that the number of contact infections is reduced quickly when the immunization rate increases.
	Moreover, we showed that combining immunization and the minimization of distinct contacts can lead to a further reduction in infections.
	
	Our paper has thus shown that simulation-optimization approaches form a promising direction for method development in network immunization problems, especially when the desire is to evaluate strategies under a rich set of conditions, such as the limited quarantining of individuals. 
	\revised{Our methods particularly apply to settings with a medium to large number of scheduled activities, such as universities or workplaces, that cover a well-defined subset of a larger population.
	Future research could focus on how to scale the developed methods to immunization at more macroscopic scales, such as at a city or national level, which have significantly larger populations and generally more limited knowledge on which contacts take place within the population.
	In this context, future research could, e.g.,} investigate the best generation of infection forests and representation of possible infection chains in the proposed stochastic programming heuristic.
	
	\revised{Another direction for future research lies in finding} robust strategies for \revised{network immunization.
	For example,} under the assumption that only a limited percentage of invited individuals will actually show up for vaccination.
	\revised{Moreover, future research could extend our stochastic programming approach to a setting in which vaccination is only partially effective.
	While the latter can be easily integrated in the genetic algorithm, it adds another layer of stochasticity in the IP model that we consider in our stochastic programming heuristic.
	Future research could, e.g., extend our sample average approximation approach or propose an L-shaped based approach to combine the idea of immunization forests with the assumption of a partially effective vaccine.
	Finally, extending our models} to sub-group selection rather than the selection of individuals could be interesting in  the considered context of contact graphs that result from planned activities.

	\begin{SecondRevised}
		\section*{Data Availability}
		The three DTU contact graphs used in this study, which are derived from the dataset as introduced by \textcite{bagger2022reducing}, are available at the following data repository: \url{https://doi.org/10.11583/DTU.30830963}. Moreover, the numerical results behind the main figures in the paper can be found at: \url{https://doi.org/10.11583/DTU.30834629}.
	\end{SecondRevised}

	\section*{Acknowledgement}
	
	This research was funded by DFF (Independent Research Fund Denmark) as part of the project \emph{FIND: Finding the ``new normal'', the power of distinct contacts (Grant number 0213-00040B)}.
	Moreover, we would like to thank Guðmundur Óskar Halldórsson and Rakel Guðrún Óladóttir, whose Master thesis provided a starting point for this paper.

	\printbibliography
	
	\appendix
	
	\pagenumbering{gobble}
	
	\begin{Revised}
		\section{\revised{Robustness Comparison Figures for a Comparison to the Random Benchmark Method}}
		\label{sec: extra robustness figures}
		
		This appendix contains additional comparison plots that supplement the analysis made in \autoref{sec: robustness}.
		In these additional plots, the Stochastic Programming Heuristic and GA are compared to the Random benchmark method.
		In particular, the relative improvement of the Stochastic Programming Heuristic and Genetic Algorithm is shown in \autoref{fig: robustness stochastic vs Random} and \autoref{fig: robustness GA vs Random}, respectively.
		
		\begin{figure}[htbp!]		
			\centering
			\includegraphics[scale=0.65]{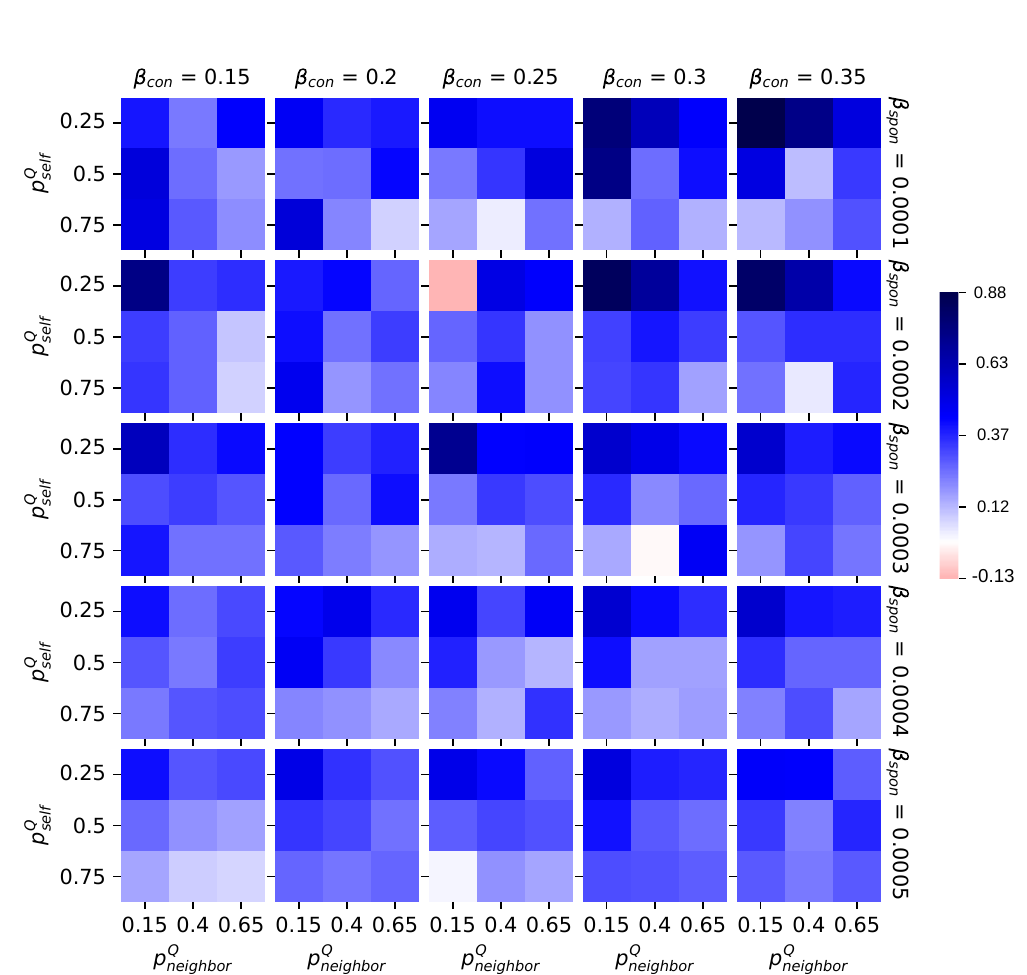}
			
			\caption{\revised{Four-dimensional plot of the relative difference in the number of contact infections between the solutions of the stochastic programming heuristic and Random benchmark method for different disease parameter configurations.}}
			\label{fig: robustness stochastic vs Random}
		\end{figure}
		
		\begin{figure}[htbp!]
			\centering
			\includegraphics[scale=0.65]{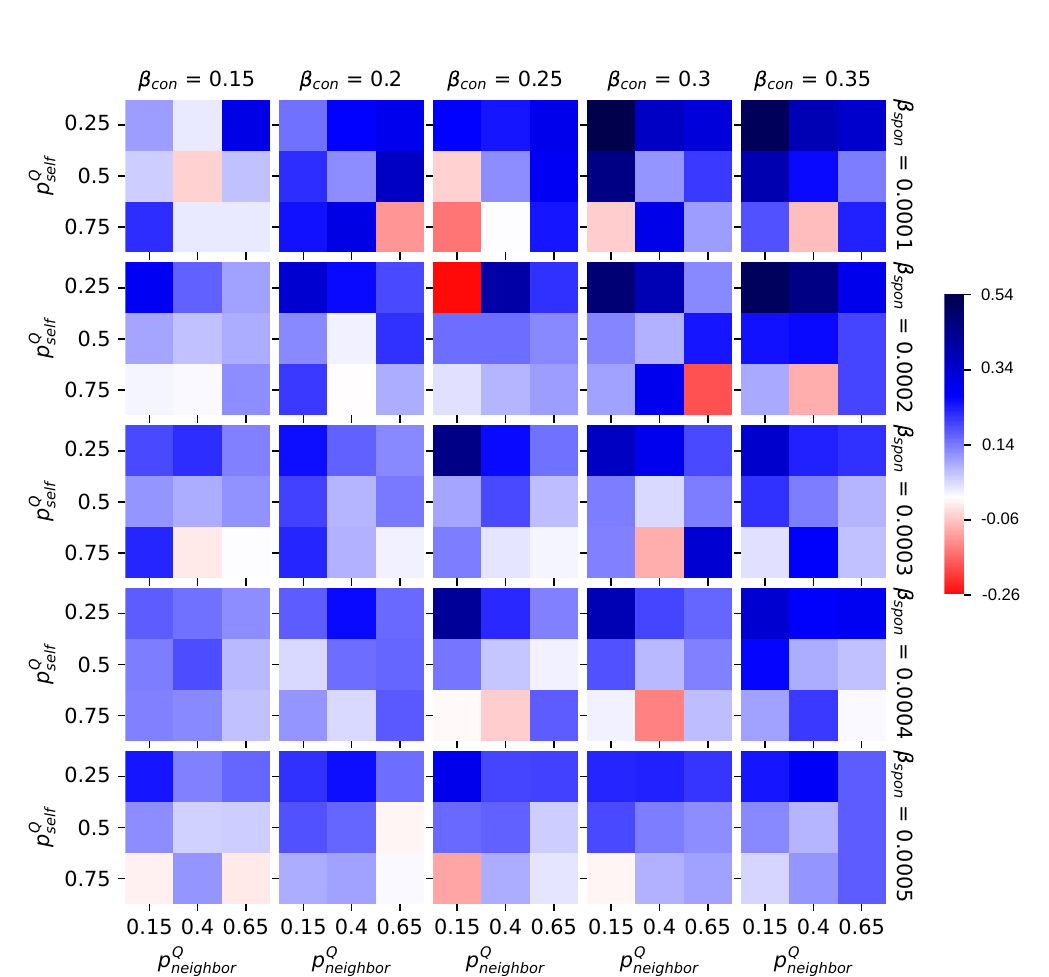}
			
			\caption{\revised{Four-dimensional plot of the relative difference in the number of contact infections between the solutions of the GA and Random benchmark method for different disease parameter configurations.}}
			\label{fig: robustness GA vs Random}
		\end{figure}
	
		\section{\revised{Statistical Analysis of Comparison to Minimizing Distinct Contacts}}
		\label{sec: stat distinct contacts}
		
		In this appendix, we analyze the statistical significance of all pair-wise comparisons between the 12 immunization and scheduling approaches as shown in \autoref{fig: immunization vs distinct contacts}.
		Here, we used the Wilcoxon signed rank test to make a comparison between the paired results of the different immunization and scheduling approaches, where instances with the same seed and thus set of starting infected nodes are paired up in the test between any two approaches.
		The results of our experiments are visualized in \autoref{fig: stat significance}.
		The figure shows that all comparisons are statistically significant at the 1\% level.
		
		\begin{figure}
			\resizebox{\textwidth}{!}{
				\input{plot.pgf}
			}
			
			\caption{\revised{P-values of the Wilcoxon signed rank test for the pair-wise comparison of the immunization and scheduling approaches shown in \autoref{fig: immunization vs distinct contacts}. Each box gives the p-value between the approach shown in the corresponding row and column.}}
			\label{fig: stat significance}
		\end{figure}
	\end{Revised}

\end{document}